\documentclass[a4paper,fleqn,usenatbib]{mnras}

\usepackage[T1]{fontenc}
\usepackage{ae,aecompl}
\usepackage{hyperref}
\hypersetup{colorlinks,linkcolor={blue},citecolor={blue},urlcolor={blue}}  
\usepackage{graphicx}
\usepackage[bigfiles]{media9}

\usepackage{graphicx}	
\usepackage{amsmath}	
\usepackage{amssymb}

\usepackage{graphicx}	
\usepackage{amsmath}	
\usepackage{amssymb}	
\usepackage{multicol}        
\usepackage{bm}		
\usepackage{pdflscape}	
\usepackage{hyperref}
\usepackage{url}
\usepackage{graphicx,times}
\usepackage{amsmath}
\usepackage[T1]{fontenc}
\usepackage{aecompl}
\usepackage{caption}
\newcommand{\x}[1]{\text{#1}}
\usepackage{booktabs,caption,fixltx2e}
\usepackage[flushleft]{threeparttable}
\usepackage{tabularx}
\usepackage{pdflscape}
\usepackage{rotating}

\usepackage[export]{adjustbox}
\usepackage{graphics}
\usepackage{amssymb,amsmath}
\usepackage{url}
\usepackage{textcomp}
\usepackage{color}
\usepackage{subcaption}
\usepackage{longtable}
\usepackage{graphicx}

\usepackage{lscape}
\usepackage{float}
\usepackage{rotating}
\usepackage{perpage}
\definecolor{ao(english)}{rgb}{0.09, 0.63, 0.13}

\title[]{WALLABY Early Science - III. An H{\sc{i}} Study of the Spiral Galaxy \mbox{NGC 1566}}

\author[A.~Elagali et al.]
{\parbox{\textwidth}{A.~Elagali$^{1,2,3}$\thanks{E-mail  ahmed.elagali@icrar.org}, L.~Staveley-Smith$^{1,2}$, J.~Rhee$^{1,2}$, O.I.~Wong$^{1,2}$,
A.~Bosma$^{4}$, T.~Westmeier$^{1,2}$, B.S.~Koribalski$^{3,2}$, G.~Heald$^{5,2}$, B.-Q.~For$^{1,2}$, D.~Kleiner$^{6,3}$, K.~Lee-Waddell$^{3}$, J.P.~Madrid$^{3}$, 
A.~Popping$^{1}$, T.N.~Reynolds$^{1,2,3}$, M.J.~Meyer$^{1,2}$, J.R.~Allison$^{2,8}$, C.D.P.~Lagos$^{1,2}$, 
M.A.~Voronkov$^{3}$, P.~Serra$^{6}$, L.~Shao$^{9,10}$, J.~Wang$^{9}$, C.S.~Anderson$^{5}$, J. D.~Bunton$^{3}$, G.~Bekiaris$^{3}$,
P.~Kamphuis$^{11}$, S-H. Oh$^{12}$, W.M.~Walsh$^{13}$, V. A. Kilborn$^{2,7}$} \vspace{0.4cm}\\ 
\parbox{\textwidth}{
$^{1}$International Centre for Radio Astronomy Research (ICRAR), M468, The University of Western Australia, 35 Stirling Highway, Crawley, WA 6009, Australia\\
$^{2}$ARC Centre of Excellence for All Sky Astrophysics in 3 Dimensions (ASTRO 3D)\\
$^{3}$Australia Telescope National Facility, CSIRO Astronomy and Space Science, P.O. Box 76, Epping, NSW 1710, Australia\\
$^{4}$Aix Marseille Univ, CNRS, CNES, LAM, Marseille, France\\
$^{5}$CSIRO Astronomy and Space Science, PO Box 1130, Bentley WA 6102, Australia\\
$^{6}$INAF-Osservatorio Astronomico di Cagliari, Via della Scienza 5, I-09047 Selargius (CA), Italy\\
$^{7}$Centre for Astrophysics \& Supercomputing, Swinburne University of Technology, PO Box 218, Hawthorn, VIC 3122, Australia\\
$^{8}$Sub-Dept. of Astrophysics, Department of Physics, University of Oxford, Denys Wilkinson Building, Keble Rd., Oxford, OX1 3RH, UK\\
$^{9}$Kavli Institute for Astronomy and Astrophysics, Peking University, Beijing 100871, China\\
$^{10}$Research School of Astronomy and Astrophysics, Australian National University, Canberra, ACT 2611, Australia\\
$^{11}$Astronomisches Institut, Ruhr-Universit$\ddot{\x{a}}$t Bochum, Universit$\ddot{\x{a}}$tsstrasse 150, 44801 Bochum, Germany\\
$^{12}$Department of Physics and Astronomy, Sejong University, 209 Neungdong-ro, Gwangjin-gu, Seoul, Republic of Korea\\
$^{13}$Solar Energy Research Institute of Singapore, National University of Singapore, Singapore 117574, Singapore \\
}}
\begin{document}

\date{Accepted 00. Received 00; in original form 00}

\pagerange{\pageref{firstpage}--\pageref{lastpage}} \pubyear{2017}

\maketitle

\label{firstpage}

\begin{abstract}
This paper reports on the atomic hydrogen gas (H{\sc{i}}) observations of the spiral galaxy
\mbox{NGC 1566} using the newly commissioned Australian Square Kilometre Array Pathfinder (ASKAP) radio telescope.
We measure an integrated H{\sc{i}} flux density of $180.2\,$Jy km\,s$^{-1}$ emanating from this galaxy, which translates to an 
H{\sc{i}} mass of $1.94\times10^{10}\,M_{\odot}$ at an assumed  distance of $21.3\,$Mpc. Our observations show that 
NGC 1566 has an asymmetric and mildly warped H{\sc{i}} disc. The H{\sc{i}}-to-stellar mass 
fraction (M$_{\x{H{\sc{i}}}}$/M$_{*}$) of \mbox{NGC 1566} is $0.29$, which is  high in comparison with galaxies that 
have the same stellar mass ($10^{10.8}\,$M$_{\odot}$). We also derive the rotation curve of this galaxy to a radius of
$50\,$kpc and fit different mass models to it. The NFW, Burkert and  pseudo-isothermal dark matter halo profiles 
fit the observed rotation curve reasonably well and recover dark matter fractions of $0.62$, $0.58$ and $0.66$, respectively. 
Down to the column density sensitivity of our observations ($N_{\x{H{\sc{i}}}}\,=\,3.7\times10^{19}\,$cm$^{-2}$), we 
detect no H{\sc{i}} clouds connected to, or in the nearby vicinity of, the H{\sc{i}} disc of \mbox{NGC 1566} nor nearby interacting systems. 
We conclude that, based on a simple analytic model, ram pressure interactions with the IGM can affect
the H{\sc{i}} disc of \mbox{NGC 1566} and is possibly the reason for the asymmetries seen in the H{\sc{i}} morphology of NGC 1566.

\end{abstract}


\section{Introduction}

\begin{figure*}
    \centering
   \includegraphics[width=.46\textwidth]{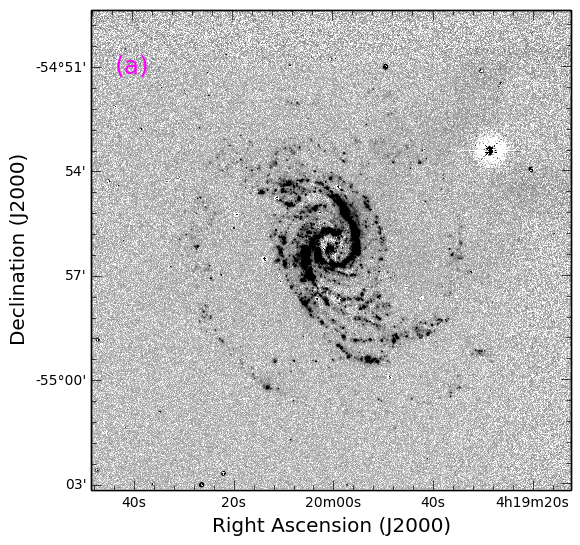}\includegraphics[width=.46\textwidth]{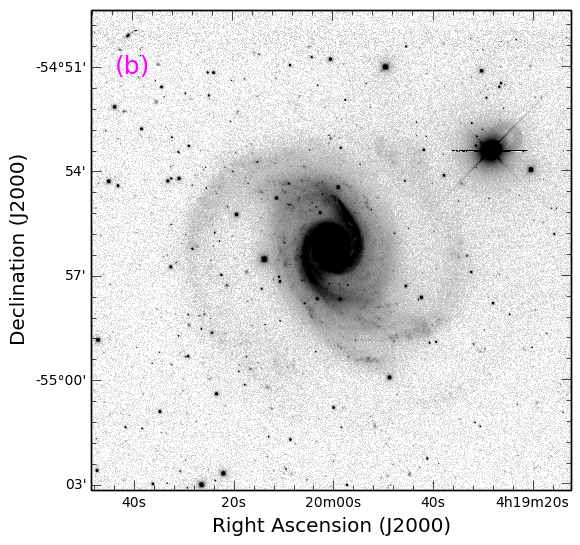}
      \includegraphics[width=0.46\textwidth]{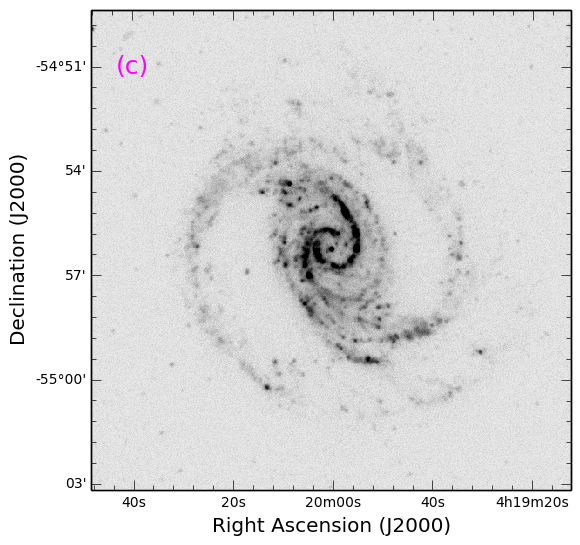}\includegraphics[width=.46\textwidth]{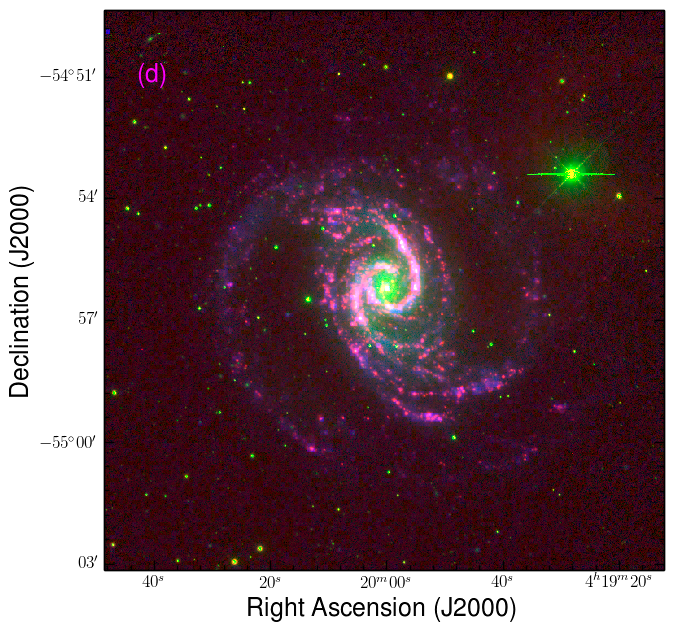}
   \caption{Panel (a) \& (b): The SINGG H$\alpha$ and $R$-band image of \mbox{NGC 1566}. Panel (c): The {\textit{GALEX}} FUV image of \mbox{NGC 1566}.
   Panel (d): The three-color RGB image of \mbox{NGC 1566}; Red represents the H$\alpha$ image, Green the $R$-band image and 
   Blue the FUV image. The H$\alpha$, and $R$-band images are convolved to $6''$, to 
   match the resolution of the FUV image \citep{2007Wong,2006Meurer}.}   
  \label{Fig1_1}
\end{figure*}
The formation and evolution of a galaxy is strongly connected to its interactions with the surrounding
local environment \citep{1986Gavazzi, 1999Okamoto, 1991Barnes, 2002Antonuccio, Balogh, 2005Avila,2005Blanton,2007Hahn,2009Fakhouri,2013Cibinel,2017Chen,2017Zheng}. The term environment is broad and not only
refers to the neighbouring galaxies but also to the tenuous gas and other material between these systems, the so-called intergalactic medium (IGM).
These interactions include gas accretion \citep{1972Larson, 1980Larson, 1988Tosi, 2008Sancisi, 2014deBlok, 2018Vulcani, 2018Rahmani},
ram pressure stripping due to the interaction with the IGM \citep{1991Dickey,1994Balsara,1999Abadi, 2006Boselli,Westmeier-2011, 2017Yoon,2018Jaff,2018Ramos}, or mergers 
and tidal harassments \citep{1972Toomre, 1982Farouki, 1996Dubinski, 2000Alonso-Herrero, Alonso2004,2009Brandl, 2009Blanton, 2015Moreno, 2018Lagos,2018Elagali,Elagali2018b}. 
There is a plethora of observational evidence on the relationship between galaxies and their local surrounding environment.
For instance, the morphology-density relationship \citep{1974Oemler, 1980Dressler,1997ADressler, 2005Postman, 2014Fogarty, 2015Houghton},
according to which early-type galaxies (ellipticals and lenticulars) are preferentially located in dense
environments such as massive groups and clusters of galaxies, whereas late-type galaxies (spirals, irregulars and/or dwarf irregulars) 
are located in less dense environments, e.g., loose groups and voids. Furthermore, galaxies in dense environments are commonly H{\sc{i}} deficient in comparison with their field counterparts
\citep{1973Davies, 1986Haynes, 1988Magri, 1990Cayatte, 2000Quilis, 2001Drinkwater, 2001Solanes,2005Omar,Sengupta2006,2007Sengupta,Kilborn2009,Chung-2009,Cortese-2011,2016denes,2017Yoon,2017Brown,2018Jung}. 
This is because dense environments promote all sorts of galaxy-galaxy and/or galaxy-IGM interactions, which have various consequences on the participant galaxies including their colour and luminosity \citep{1992Loveday, 2001Norberg, 2005Croton,2012Kreckel,2014McNaught} 
as well as their star formation rates \citep{1997Balogh, 1999Poggianti, 2002Lewis, 2003Gomez,2004Kauffmann,2008Porter, 2018Paulino, 2018Xie}. Hence, a complete picture of the local environments surrounding 
galaxies is essential to distinguish between the different mechanisms that enable galaxies to grow, or quench \citep{2010Thomas, 2010Peng,2012Putman,2016Marasco,2019Bahe}.\\

The effects of galaxy interactions with the surrounding environment are most noticeable in the outer
discs of the participant galaxies. However, H{\sc{i}} discs are
better tracers of the interactions with the local environment, because H{\sc{i}} discs are commonly more extended
and diffuse than the optical discs and, as a consequence, more sensitive to external influences
and susceptible to hydrodynamic processes unlike stars \citep{Yun-1994,Braun-2004,Michel-Dansac-2010}.
Probing the faint H{\sc{i}} gaseous structures in galaxies is an onerous endeavour as it requires high observational sensitivity to low surface brightness features,
which means long integration times. Such observations are now possible on unprecedentedly 
large areas of the sky. These wide field H{\sc{i}} surveys  will be conducted using state-of-the-art radio telescopes that have subarcminute angular 
resolution, for instance the Australian  Square Kilometre Array Pathfinder \citep[ASKAP;][]{2007Johnston, 2008Johnston}, the Karoo Array
Telescope \citep[MeerKAT;][]{2016Jonas} as well as the APERture Tile In Focus \citep[APERTIF;][]{2008Verheijen}. \\

The Widefield ASKAP L-band Legacy All-sky Blind surveY (WALLABY) is an H{\sc{i}} imaging survey that will be 
carried out using ASKAP radio telescope to image $3/4$ of the sky out to a redshift of $z \sim 0.26$ 
\citep{2012Koribalski}. ASKAP is equipped with phased-array feeds \citep[PAFs,][]{2008Hay}, which can deliver a
field-of-view of $30\,$ square degrees, formed using $36$ beams at $1.4\,$GHz \citep{2012Koribalski}.
WALLABY will provide H{\sc{i}} line cubes with relatively high sensitivity to diffuse emission with a root-mean-square (rms) noise of 
$1.7\,$mJy beam$^{-1}$ per $4$ km s$^{-1}$ channel and will map at least $500,000$ H{\sc{i}} emitting galaxies over its
entire volume \citep{2012Duffy}.
Hence, WALLABY will help revolutionise our understanding of the behaviour of the  H{\sc{i}} gas in different environments  and the  distribution
of  gas-rich  galaxies in their local environments.
The prime goal of this work is to provide the community with an example of the capabilities of the widefield H{\sc{i}} spectral 
line imaging of ASKAP and the scope of the science questions that will be addressable with WALLABY. 
We present the ASKAP H{\sc{i}} line observations of the spiral galaxy \mbox{NGC 1566} (also known as WALLABY J041957-545613) and validate these with archival single-dish and interferometric H{\sc{i}}
observations from the Parkes $64\,$m telescope using the $21$-cm multibeam receiver \citep{1996Staveley} and the Australia Telescope 
Compact Array (ATCA), respectively. Further, we use the sensitivity and angular resolution of the ASKAP H{\sc{i}} observations to study the
asymmetric/lopsided H{\sc{i}} gas morphology and warped disc of \mbox{NGC 1566} and attempt to disentangle the different environmental processes affecting 
the gas and the kinematics of this spectacular system. \mbox{NGC 1566} is a face-on SAB(s)bc spiral galaxy \citep{1973Vaucouleurs, 1976Vaucouleurs}, and is part of 
the Dorado loose galaxy group \citep{1995Bajaja,2004Aguero,2005Kilborn}.\\

\begin{table}
\centering
\caption{The properties of \mbox{NGC 1566} adopted from the literature and reported in this paper.}
\label{my-label}
\begin{tabular}{lll}\\ \hline \hline
Property& \mbox{NGC 1566}& Reference \\ \hline \hline
Right ascension (J2000) &04:20:00.42&\citet{1973Vaucouleurs} \\
Declination (J2000) &$-$54:56:16.1& \citet{1973Vaucouleurs}\\
Morphology &SAB(rs)bc&\citet{1973Vaucouleurs} \\
H{\sc{i}} systemic velocity (km\,s$^{-1}$)   &   $1496$ & This work    \\
$L_{\x{R}}$ (L$_{\odot}$) &  $1.2\times10^{11}$  & \citet{2006Meurer}\\
$L_{\x{FIR}}$ (L$_{\odot}$) &  $2.5\times10^{10}$& \citet{2003Sanders}  \\
$\mu_{R}$ (ABmag arcsec$^{-2}$) & $19.3$& \citet{2006Meurer}\\
$\x{m}_{I}$ (ABmag) & $8.7$& \citet{1997Walsh}\\
$\x{m}_{3.6\mu m}$ (ABmag) & $10.15$& \citet{2014Laine}\\
$\x{m}_{V}$ (ABmag) & $9.8$& \citet{1997Walsh}\\
M$_{*}$ (M$_{\odot}$) & $6.5\times10^{10}$& This work\\
H$\alpha$ Equivalent width (\AA{}) & $39\pm3$ & \citet{2006Meurer}\\
M$_{\x{FUV}}$ (ABmag) & $-20.66$ &\citet{2007Wong}\\
M$_{\x{H{\sc{i}}}}$ (M$_{\odot}$) &  $1.94\times10^{10}$& This work\\
M$_{\x{H2}}$ (M$_{\odot}$) & $1.3\times10^{9}$& \citet{1995Bajaja} \\
M$_{\x{BH}}$ (M$_{\odot}$) & $8.3\times10^{6}$& \citet{2002Woo} \\
Position angle (degree) & $219\pm4$& This work \\
Inclination (degree) & $31\pm7$& This work \\
SFR$_{\x{H}\alpha}$(M$_{\odot}$ yr$^{-1}$) & $21.5$&\citet{2006Meurer} \\
D$_{25}$ (kpc)  & $35$  &\citet{1997Walsh}\\
Distance (Mpc) & $21.3$ & \citet{2005Kilborn}\\ \hline
\end{tabular}
\end{table}

Figure \ref{Fig1_1} shows the Survey for Ionisation in Neutral Gas Galaxies \citep[SINGG;][]{2006Meurer} H$\alpha$
and $R$-band images, the {\textit{GALEX}} FUV image and a three-color RGB image of \mbox{NGC 1566}, where the H$\alpha$, $R$-band, and FUV
images present the Red, Green and the Blue, respectively. This galaxy has a weak central bar  \citep[north-south orientation 
with a length of $\sim32''.5$,][]{1983Hackwell} and two prominent star-forming spiral arms that form a pseudo-ring
in the outskirts of the optical disc \citep{1982Comte}. The H$\alpha$ emission map of \mbox{NGC 1566} is dominated by a small but extremely bright H$\alpha$ complex region
located in the northern arm that emits approximately a quarter of the total disc's  H$\alpha$ flux when excluding
the emission from the nucleus \citep{1990Pence}. \mbox{NGC 1566} hosts a low-luminosity Seyfert nucleus  \citep{2002Reunanen, 2009Levenson, 2014Combes, 2017daSilva} known for its
variability from the X-rays to IR bands \citep{1986Alloin, 2004Glass}. The origin of the variability in AGNs is still controversial
and is hypothesised to be caused by processes such as accretion prompted by disc instabilities, surface temperature fluctuations, or even variable heating from coronal X-rays 
\citep{1986Abramowicz, 1993Rokaki, 2012Zuo, 2014Ruan,2016Koz}. Although, \mbox{NGC 1566} has been subjected
to numerous detailed multiwavelength studies, as cited above, this work presents the first detailed H{\sc{i}}
study of this spiral galaxy. \mbox{Table 1 presents} a summary of the relevant properties of \mbox{NGC 1566} from the literature and 
from this paper.\\

\begin{table*}
\begin{tabular}{llllll} 
  & \multicolumn{5}{c}{{\bf{Table 2.}} WALLABY early science observations: Dorado field}  \\ \cline{2-4} \hline \hline
 Observation Dates & Time on Source (hrs) &  Bandwidth (MHz)& Central Frequency (MHz)& Number of Antennas & Footprint \\ \hline
 28 Dec 2016 & 11.1  &   &   &  10  & A \\
 29 Dec 2016 &11.2   &  &  &10  &B\\
 30 Dec 2016 & 11.1  &  &  &9  &A\\
 31 Dec 2016 & 12.6 & 192 & 1344.5 &10  &B\\
 01 Jan 2017 &  11.1 &  &  & 10 &A\\
 02 Jan 2017 &12.0  &  &  & 10 &B\\
 03 Jan 2017 &3.8  &  &  & 9 &A\\ \hline
  23 Sep 2017 &12.0  &  &  & 12 &A\\
 24 Sep 2017 & 12.0 &240  & 1368.5  & 12&B\\
 25 Sep 2017  &  4.0 &  &  & 12&A\\
 26 Sep 2017 & 12.0 &  & &12 &B\\ \hline
 27 Sep 2017 &12.0  &  & &12 &A\\
 28 Sep 2017 & 12.0 &  & &12 &B\\
 15 Dec 2017 &9.1  & 240 & 1320.5 &16 &A\\
 03 Jan 2018 & 12.0 &  &  & 16 &B\\
 04 Jan 2018 & 9.0 &   &   & 16&A\\ \hline
\end{tabular}
\end{table*}

This paper is organised as follows: in Section 2, we describe the WALLABY early science observations and reduction procedures along with the previously
unpublished archival data obtained from the ATCA online archive. Section 3 describes our main results, 
in which we provide detailed analysis of the H{\sc{i}} morphology and kinematics of this galaxy.
In Section 4, we fit the observed rotation curve to three different 
dark matter halo models, namely, the pseudo-isothermal, the Burkert and the Navarro-Frenk-White (NFW) halo profiles. Section 5 discusses the possible scenarios 
leading to the asymmetry in the outer gaseous disc of \mbox{NGC 1566} and presents evidence that ram pressure could  be the main cause of this asymmetry.
In Section 6, we summarise our main findings. For consistency with previous H{\sc{i}} and X-ray studies of the \mbox{NGC 1566} galaxy group, we adopt a distance of
$21.3\,$Mpc \citep{2005Kilborn, 2004Osmond}, which is based on a Hubble constant of $H_{0}\,=\,70.3\,$km\,s$^{-1}$~Mpc$^{-1}$, 
though we note that this is at the upper end of values quoted in the NASA/IPAC Extragalactic Database (NED). For more details
refer to Section 4.5. 
\begin{figure}
    \centering
    \includegraphics[width=0.48\textwidth]{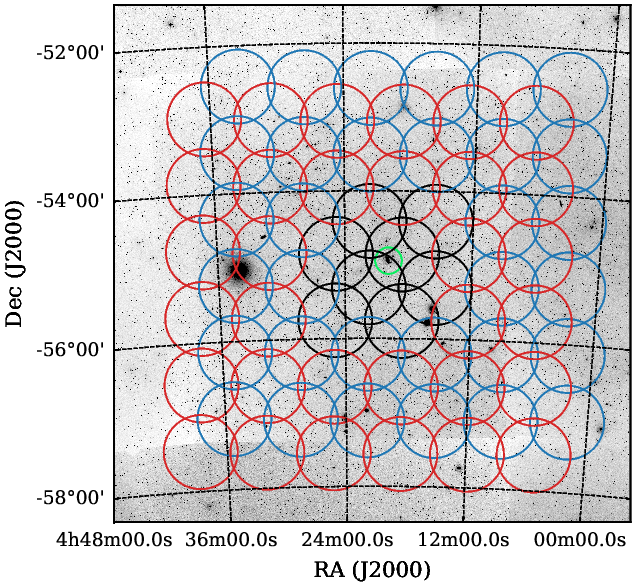}
    \caption{The Dorado observations footprint, where blue and red circles represent footprints A and B, respectively. 
    The inner eight black circles represent the beams around \mbox{NGC 1566} used in this paper.
    The background image is the blue band Digitised Sky Survey (DSS) image, and the green circle highlights the location of \mbox{NGC 1566}.}
    \label{fig2_1}
\end{figure}

\section{Data}
\subsection{WALLABY Early Science Observations}
ASKAP is situated in the Murchison Radioastronomy Observatory in Western Australia,
a remote radio-quiet region about $305\,$km North-East of Geraldton in Western Australia. ASKAP is one of the new generation of radio 
telescopes designed to pave the way for the Square Kilometre Array \citep[SKA;][]{2009Dewdney}. This SKA precursor consists of $36$
separate $12$-metre radio dishes that are located at 
longitude $116.5^{\circ}$ east and latitude $26.7^{\circ}$ south\footnote{\url{https://www.atnf.csiro.au/projects/askap/index.html}}. Each 12-metre antenna has a single reflector on an azimuth-elevation 
drive along with a third axis (roll-axis) to provide all-sky coverage and an antenna surface capable of operation up to $10\,$GHz.
The antennas are equipped with Mark two (MK{\,\sc{ii}}) phased array feeds (PAFs), which 
provide the antennas with a $30$ square degree field-of-view, making this radio telescope a surveying machine
\citep{2009DeBoer,2016Schinckel}. During the first five years of operation, ASKAP will mainly carry out observations
for ten science projects, one of which is WALLABY\footnote{\url{https://wallaby-survey.org/}}.\\


\begin{figure}
    \centering
   \includegraphics[width=0.475\textwidth]{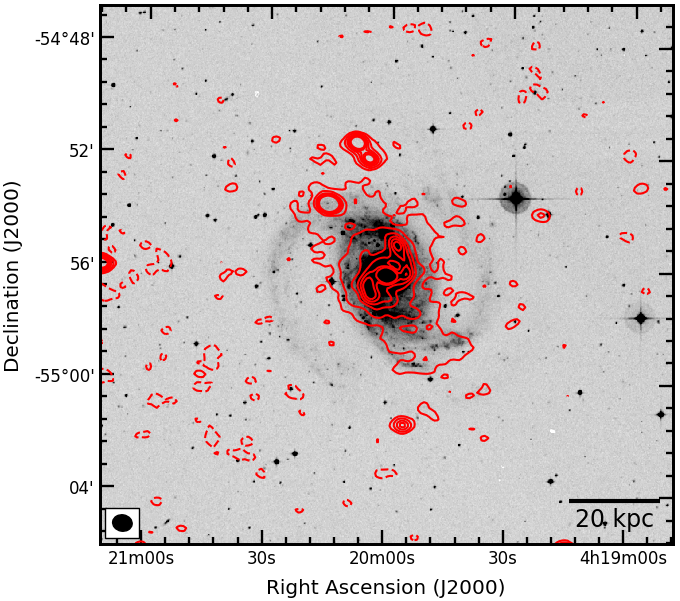}
    \caption{Optical DSS image (blue band) on a linear stretch with contours from
    ASKAP continuum map overlaid at $-2,2,5,10,15,20,25$ multiplied by $1\,\sigma$ ($0.3\,$mJy beam$^{-1}$). The $-2\,\sigma$
level is displayed with dashed contours. The restored beam of ASKAP observation is in the bottom left.} 
    \label{fig2_3}
\end{figure}

In October 2016, ASKAP early science observations program started using $12$ PAF-equipped ASKAP antennas 
(out of $36$ antennas) to pave the way and improve the data reduction and analysis techniques while commissioning ASKAP to full
specifications. Over $700\,$ hours of early science observations  were dedicated to WALLABY, during which four fields were observed 
to the full survey sensitivity depth, rms noise sensitivity per $4$ km s$^{-1}$ channel  of $1.7\,$mJy beam$^{-1}$ 
\citep{karen2019, reynolds}. One of these fields is the Dorado early science field. Figure \ref{fig2_1} shows the two interleaves of the 
square $6\times6$ beam footprint of the observations overlaid on the Digitised Sky Survey (DSS) image of this field. The blue and red 
beams are referred to as footprint A (centred at$\,\alpha\,=\,04$:$18$:$35$,$\,\delta\,=\,-54$:$51$:$43\,$; J$2000$) and B 
(centred at$\,\alpha\,=\,04$:$21$:$44.7$,$\,\delta\,=\,-55$:$18$:$33.8\,$; J$2000$), respectively.
Mosaicking the two interleaves reduces the noise level at the edge of the beams and produces a smoother noise pattern.
The inner eight black circles represent the beams around \mbox{NGC 1566} used  to make the H{\sc{i}} cube
that forms the basis of this work. The green circle shows the location of \mbox{NGC 1566}.
The observations of this field started in December $2016$ and were completed in January 2018 using at the beginning 
only $9\,$-$\,10$ ASKAP antennas, however the most recent observations were completed using $16$ ASKAP
antennas. The baseline range of the ASKAP array for the current observations is between $22$-$2305\,$m,
which ensured excellent \textit{uv}-coverage and a compromise between the angular resolution and the surface brightness sensitivity of the observations. 
The bandwidth of the Dorado observations ranges between $192$ and $240\,$MHz, due to upgrades to the correlator capacity during the early science period, and has a channel width of $18.5\,$kHz ($4\,$km\,s$^{-1}$).
For each day of observation, the primary calibrator, PKS1934-638, is observed at the beginning for two to three hours and is positioned at the
centre of each of the $36$ beams. The total on-source integration time is $167.0\,$hr; $83.2\,$hr for footprint A and $83.8\,$hr for footprint B. Refer to Table 2 
for a summary of the Dorado early science observations.\\
\begin{table*}
\begin{center}
\label{table1}
\begin{tabular}{lll}
&\multicolumn{1}{l}{{\bf{Table 3.}} ASKAP and ATCA H{\sc{i}} Observations Results} \\ \hline \hline
Parameter& ASKAP value & ATCA value \\ \hline
rms noise (Jy beam$^{-1}$ per $4\,$km\,s$^{-1}$ channel )& $1.7\times10^{-3}$&$2.4\times10^{-3}$\\
Synthesised Beam Size (arcsec$\times$arcsec)&$42\times35$&$45\times40$\\
Synthesised Beam Size (kpc$\times$kpc)&$4.3\times3.6$&$4.6\times4.1$\\
Beam PA (degree)&$80.0$&$6$\\
Channel width (km\,s$^{-1}$)&$4$&$4$\\ 
Channel map pixel size (arcsec$\times$arcsec)&$6\times6$&$6\times6$\\
\mbox{NGC 1566} H{\sc{i}} total flux (Jy km\,s$^{-1}$) & $180.2\pm16.3$&$177.0\pm14.0$\\ 
\mbox{NGC 1566} peak flux density (Jy) & $1.35\pm0.09$&$1.24\pm0.02$\\
\mbox{NGC 1566} H{\sc{i}} mass (M$_{\odot}$) & $(1.94\pm0.18)\times10^{10}$ &$(1.91\pm0.15)\times10^{10}$ \\
$w_{50}$ Line Width (km\,s$^{-1}$) & $208\pm10$&$201\pm8$ \\ \hline
\end{tabular}
\end{center}
\end{table*}

We use ASKAP{\textsc{\scriptsize{SOFT}}}\footnote{\url{https://www.atnf.csiro.au/computing/software/askapsoft/sdp/docs/current/index.html}}
to process the Dorado observations.
ASKAP{\textsc{\scriptsize{SOFT}}} is a software processing pipeline developed by the ASKAP computing team to do the calibration, spectral
line and continuum imaging, as well as the source detection for the full-scale ASKAP observations in a high-performance computing environment.
This pipeline is written using C++ and built on the casacore library among other third party libraries. A 
comprehensive  description of ASKAPsoft reduction pipeline is under preparation in Kleiner et al. (in prep.). The reader 
can also refer to \citet{karen2019} or \citet{reynolds} for a similar brief description of the reduction and pipeline procedures. For each day of observation, we flag and calibrate the measurement data set  on a per-beam basis 
using ASKAP{\textsc{\scriptsize{SOFT}}} tasks {\sc{cflag}} and {\sc{cbpcalibrator}}, respectively.
Using the {\sc{cflag}} utility, we flag the autocorrelations and the spectral channels affected by radio frequency interference (RFI) by applying
a simple flat amplitude threshold. Then, we apply a sequence of Stokes-V flagging and dynamic flagging of amplitudes, integrating over
individual spectra. We process the central $8$ beams ($4$ in each footprint) of each observation using an $8\,$MHz bandwidth ($432\,$channels), 
between $1410$ to $1418\,$MHz (velocity range between $511$ to $2196$ km s$^{-1}$) to save computing time and disc space on the Pawsey
supercomputer. We then process the calibrated visibilities to make the continuum images using the task {\sc{imager}} and self-calibrate (three loops) to 
remove any artefacts or sidelobes from the continuum images.\\

\begin{table*}
\begin{tabular}{lllll} 
  & \multicolumn{3}{c}{{\bf{Table 4.}} Archival ATCA H{\sc{i}} Observations: Instrumental Parameters}  \\ \cline{2-5} \hline \hline
 Parameter & \multicolumn{3}{c}{Array Configuration}  \\ \cline{2-5} 
 &   375B   & 1.5D    &   750B   &  1.5C \\ \hline
 Observation Dates & 1994 April 05  & 1994 May 26 & 1994 June 01  &   1994 June 17          \\
On-source integration time (hrs) & 9      &  9     &  9 & 9 \\ 
Shortest Baseline (m)&  31    &107 &  61   &   77       \\ 
Longest Baseline (m) &   5969    &     4439 &   4500   & 4500      \\ 
Central Frequency (MHz) &   1413    &  1413     & 1413  & 1413  \\ 
Bandwidth (MHz) &   8    &  8     & 8 & 8  \\ \hline 
\end{tabular}
\end{table*}

Prior to the spectral-line imaging stage, we subtract radio continuum emission from the visibility data set using the best-fit continuum sky model
produced in the previous step. Thereafter, we combine the data set for each beam, 
seven nights in footprint B and nine nights in footprint A, in the \textit{uv} domain and image using the 
ASKAP{\textsc{\scriptsize{SOFT}}} task \textsc{imager} with a robust weighting value of $+0.5$ and a $30$ arcsec Gaussian taper.
We clean the combined image for each beam using a major and minor cycle threshold of $3\sigma$, three times  the theoretical rms noise for the seven observations combined.
To measure the theoretical rms noise of each epoch, we use the  on-line sensitivity calculator
\footnote{\url{http://www.atnf.csiro.au/people/Keith.Bannister/senscalc/}}, with antenna efficiency value of $0.7$ and system
temperature of $50\,$K. For instance, the theoretical rms noise for ten hours using $10$, $12$ or $16$ 
antennas is $5.03,4.15$ and $3.08\,$mJy beam$^{-1}$, respectively. Then, we subtract the residual continuum emission from the restored
cube using the ASKAP{\textsc{\scriptsize{SOFT}}} task \textsc{imcontsub} and mosaicked the $8$ beams using the ASKAP{\textsc{\scriptsize{SOFT}} task
\textsc{linmos}}. Even though mosaicking different ASKAP beams with \textsc{linmos} can introduce 
correlated noise to the final H{\sc{i}} cube, this has no effect on the final flux scale, and only a minor effect on the rms noise.
This image domain continuum subtraction is necessary to obtain a higher dynamic range H{\sc{i}} cube
and remove any remaining residual artefacts, which aids source finding and parameterisation. 
Figure \ref{fig2_3} shows the optical DSS image (blue band) overlaid with contours from the ASKAP continuum map.
The total $21\,$cm continuum flux density of \mbox{NGC 1566} from ASKAP observations is $199\pm3\,$mJy, in agreement with the value of $204\pm28\,$mJy 
(at $21.7\,$cm ) reported by \citet{1996Ehle} using the ATCA. The restored synthesised beam has a size of 
$\theta_{\x{FWHM}}\,=\,42''\times35''$ with a position angle of $\,80^\circ$. 
The H{\sc{i}} line cube has rms noise per $4\,$km s$^{-1}$ channel of $1.7$ mJy beam$^{-1}$, which translates to 
a $3\sigma$ column density sensitivity of $N_{\x{H{\sc{i}}}}\,=\,1.5\times10^{19}\,$cm$^{-2}$.  
The rms noise in a wider channel width ($20\,$km s$^{-1}$) equals $0.84$ mJy beam$^{-1}$, and corresponds to
a $3\sigma$ column density sensitivity of $N_{\x{H{\sc{i}}}}\,=\,3.7\times10^{19}\,$cm$^{-2}$.
The properties of the final cube are summarised in Table 3.\\

\begin{figure}
    \centering
    \includegraphics[width=0.48\textwidth]{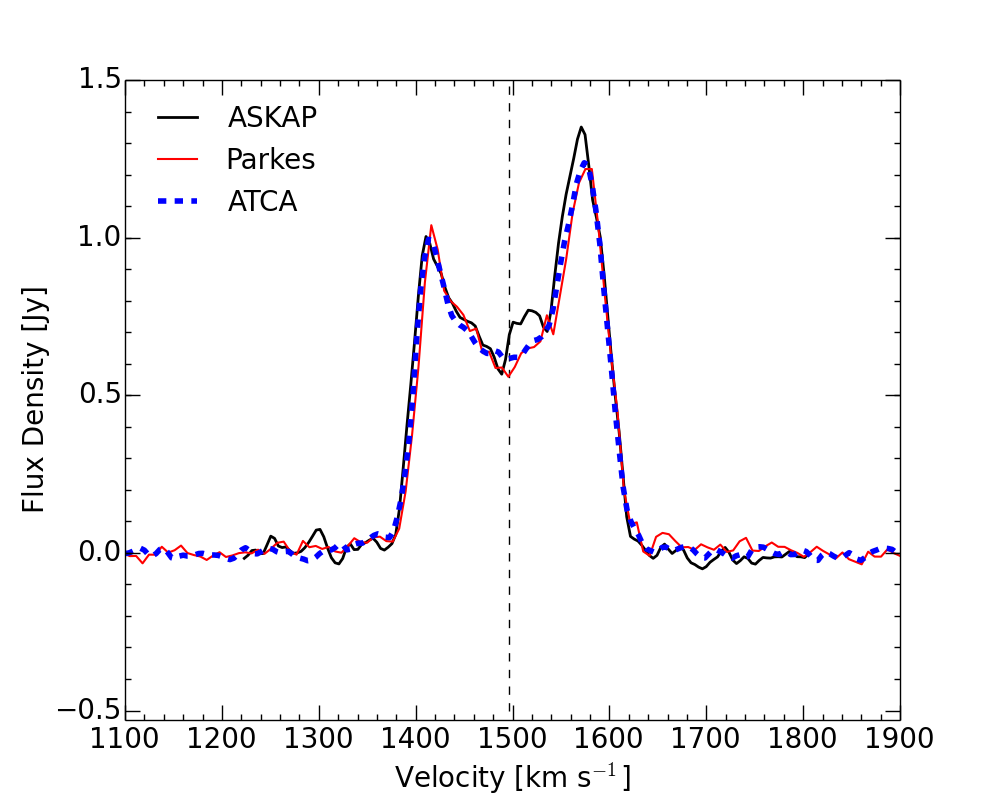}
    \caption{The H{\sc{i}} line profile of \mbox{NGC 1566} from our ASKAP \mbox{observations} (black line), from
   the ATCA (blue dashed line) and Parkes single-dish observations (red line).
    The vertical line delimits the H{\sc{i}} systemic velocity of \mbox{NGC 1566}
    $V_{\x{sys}}\,=\,1496\,$km\,s$^{-1}$ derived from our kinematics analysis (refer to Section 3).}
    \label{fig3_1}
\end{figure}

\begin{figure*}
\begin{center}
\includegraphics[width=0.8\textwidth]{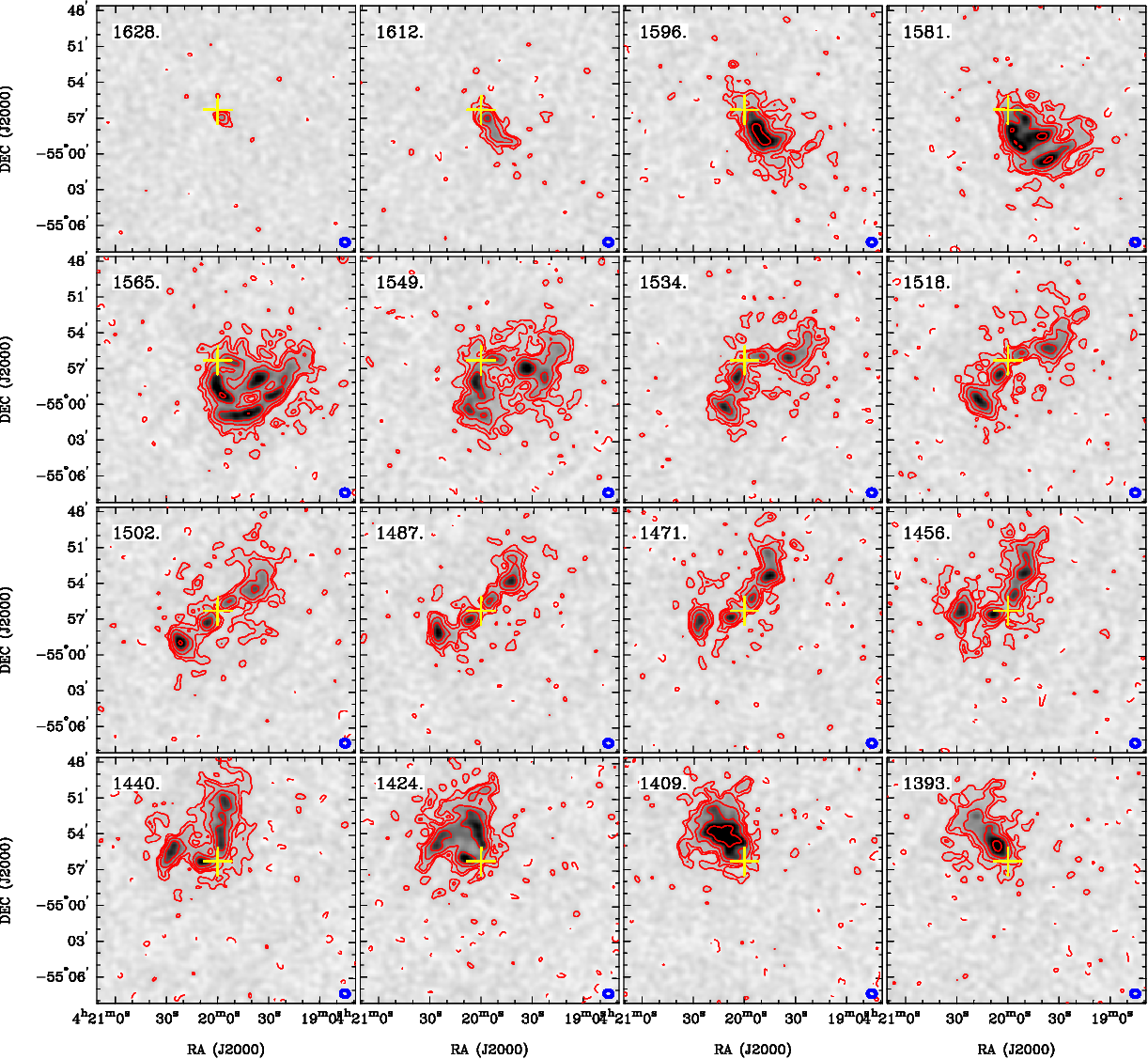}
\caption{The individual H{\sc{i}} channel maps of \mbox{NGC 1566} from ASKAP observations
in the velocity range between $1628$-$1393\,$km\,s$^{-1}$ and with a step-size  of $16\,$km\,s$^{-1}$.
The velocity of each channel is shown in the top-left, whereas the restored beam is shown in the
bottom-right (blue ellipse). Contours are at $-3,3,5,10,20,40\,$ times the $1\,\sigma$ noise ($0.85\,$ mJy beam$^{-1}$  per $16\,$km\,s$^{-1}$ channel). The yellow cross marks the kinematic centre  of \mbox{NGC 1566}
derived using {\sc{rotcur}} (refer to Section 3).}
\label{fig3_4}
\end{center}
\end{figure*}

\subsection{ATCA Observations}

\begin{figure}
\begin{center}
\includegraphics[width=0.49\textwidth]{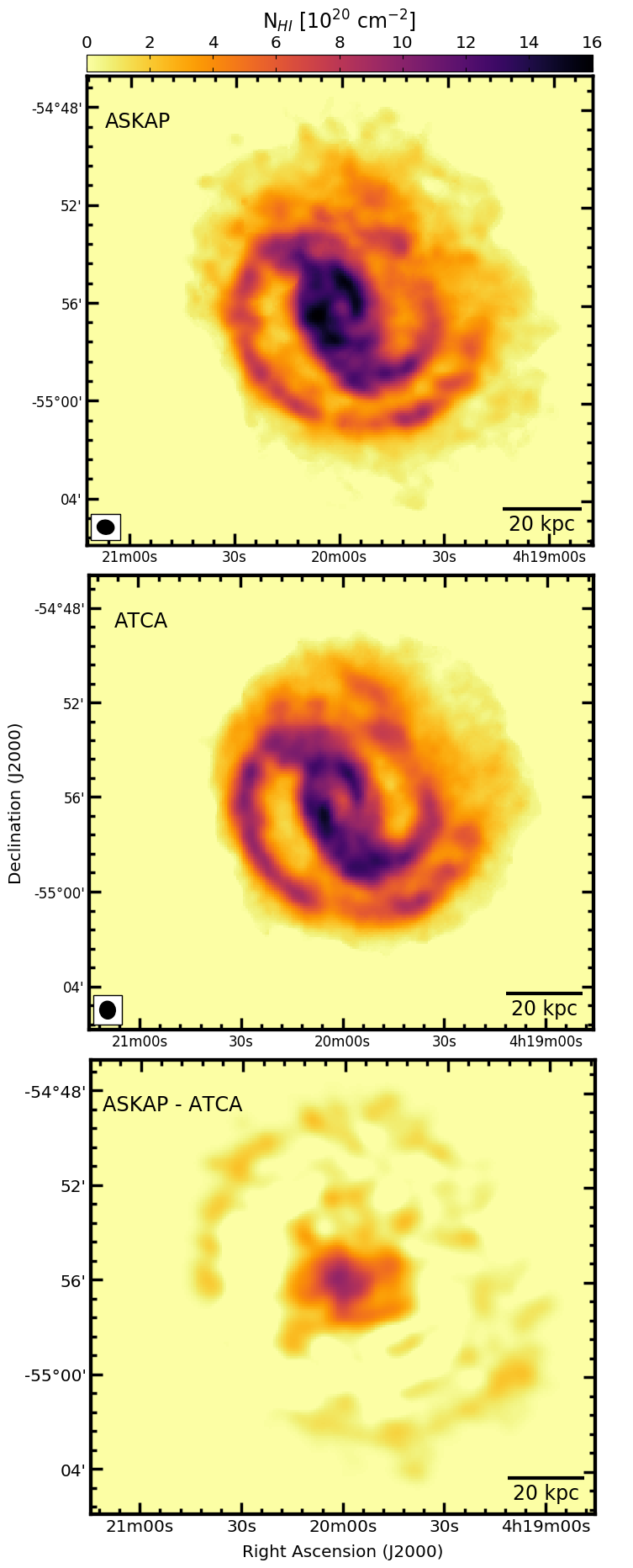}\\
\caption{The H{\sc{i}} column density map of \mbox{NGC 1566} as obtained from ASKAP observations (top panel),
the archival ATCA observations (middle panel), and a difference between the two maps (bottom panel). To produce the 
difference map, we have convolved the ASKAP and ATCA maps to the same beam angular resolution.
The ellipse shows the restored beam of ASKAP observations ($\theta_{\x{FWHM}}\,=\,42''\times35''$) and the ATCA observations 
($\theta_{\x{FWHM}}\,=\,45''\times40''$).}
\label{fig34c}
\end{center}
\end{figure}

\mbox{NGC 1566} was observed using the ATCA in four epochs between April 1994 and June 1994 with each epoch being $12\,$h in duration.
Four different ATCA array configurations were used to observe \mbox{NGC 1566}, namely the 1.5C, 375, 750B as
well as the 1.5D  configuration \citep{1997Walsh}. These configurations have baseline distances 
in the range between $31$ and $5969\,$m. In each of the four epochs, the observation was centred at $1413\,$MHz 
(the redshifted $1420\,$MHz line frequency for NGC 1566) for a duration of nine hours on the source (a total of $36\,$hrs, see Table 4).
The bandwidth of these observations is $8\,$MHz, over $512$ velocity channels. Hence, each channel corresponds to $15.625\,$kHz in width,
and a $3.3\,$km\,s$^{-1}$ velocity resolution. The ATCA primary calibrator, PKS1934$-$638 (flux density $S_{\x{1396 MHz}}=14.9\,$Jy), was observed before 
each epoch of observations for a duration of $30\,$minutes and used as the bandpass calibrator. 
The phase calibrator PKS0438$-$436 ($S_{\x{1396 MHz}}=6.09\,$Jy) was 
observed each hour during the four epochs for a duration of $15$ minutes to ensure the precision of the calibration.\\

\begin{figure*}
\begin{center}
\includegraphics[width=0.93\textwidth]{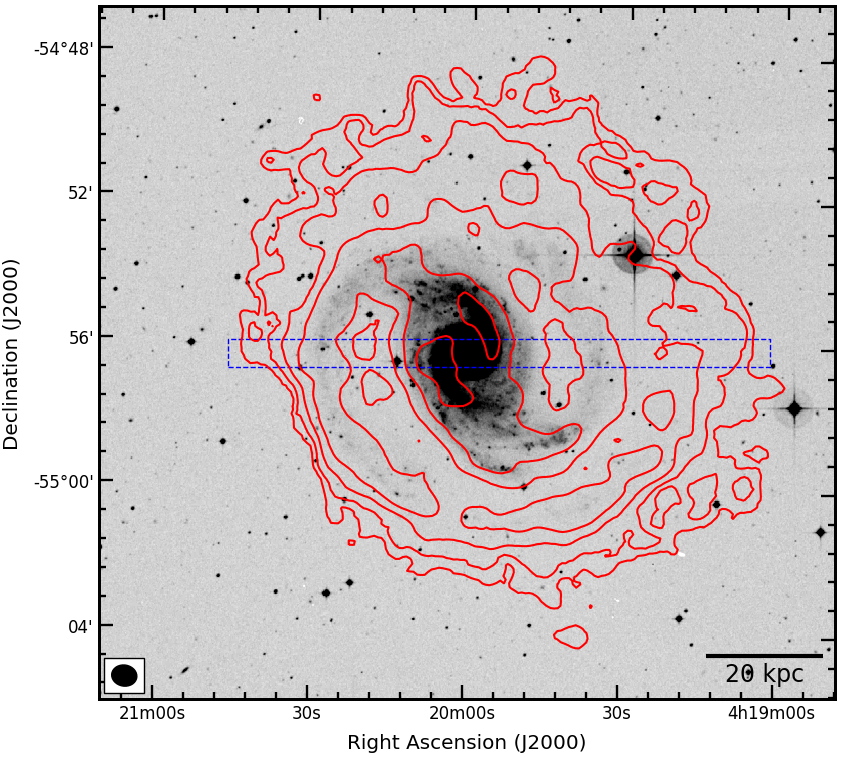}
\caption{DSS blue band image of \mbox{NGC 1566} with column density contours from ASKAP observations overlaid 
at $(0.6,1.2,2.4,4.8,9.6,14.0, 16.2)\times10^{20}\,$cm$^{-2}$. The ellipse shows the restored beam of ASKAP observations.
Left-to-right in the figure corresponds to east-to-west of the galaxy. The blue-dashed line corresponds to 
the rectangular cut through the galaxy, with a width of $5\,$kpc, that is used to measure the cut in the column density 
distribution across \mbox{NGC 1566} in Figure \ref{fig3_3}.}
\label{fig3_2}
\end{center}
\end{figure*}

We follow the standard procedures described in \citet{2018Elagali} to flag, calibrate and image the
H{\sc{i}} line observations using the \textsc{miriad} package \citep{Sault1995}.
We use the  \textsc{miriad} \textsc{uvlin} task \citep{1994Sault, Cornwell1992}
to subtract the radio continuum from the visibility data set. Then, we use \textsc{miriad} \textsc{invert} task to Fourier-transform the continuum subtracted visibilities to a map with 
robust weighting parameter of $+0.5$ and a symmetric taper of $15$ arcsec to have an  optimal sidelobe suppression and intermediate weighting between 
uniform and natural. We also re-sample to the ASKAP resolution of $4\,$km\,s$^{-1}$ at this stage, and apply 
the \textsc{clean} task down to three times  the theoretical rms noise. As a final step, we apply the primary beam correction using the \textsc{linmos} task.
The synthesised beam-size is $\theta_{\,\x{FWHM}}=45''\times40''$ with {\textit{PA}}$\,=\,6^\circ$. 
The cube has  rms noise per $4\,$km\,s$^{-1}$ channel of $2.4\,$mJy beam$^{-1}$, close to the theoretical rms noise of our observations ($2.2\,$mJy beam$^{-1}$).
The $3\sigma$ column density sensitivity per $4\,$km\,s$^{-1}$ channel is
$N_{\x{H{\sc{i}}}}\,=\,1.7\times10^{19}\,$cm$^{-2}$. Over a $20\,$km s$^{-1}$ channel width the rms noise equals
$1.2$ mJy beam$^{-1}$ and the corresponding $3\sigma$ column density sensitivity is $N_{\x{H{\sc{i}}}}\,=\,4.3\times10^{19}\,$cm$^{-2}$.\\

\section{Gas Morphology and Kinematics}

\subsection{H{\sc{i}} Morphology and distribution in \mbox{NGC 1566}}
Figure \ref{fig3_1} presents the integrated H{\sc{i}} spectrum of \mbox{NGC 1566} as obtained from ASKAP observations (black line), 
unpublished archival ATCA observations (blue dashed-line) and re-measured Parkes single-dish spectrum
(red line) from \citet{2005Kilborn}. We measure a total flux value of $180.2\pm16.3\,$Jy km s$^{-1}$ from ASKAP observations. 
This flux value corresponds to a total H{\sc{i}} mass of $1.94\times10^{10}\,$M$_{\odot}$, assuming that \mbox{NGC 1566} is at 
a distance of $21.3\,$Mpc. The integrated H{\sc{i}} flux density of \mbox{NGC 1566} from  ASKAP early science observations
is  within the  expected error of the value measured from the ATCA observations ($177.0\pm14.0\,$Jy km s$^{-1}$) and 
from Parkes observations ($175.0\pm7.4\,$Jy km s$^{-1}$).
We note that the integrated flux density of \mbox{NGC 1566} derived from our ATCA data reduction is within error of the value reported 
in the thesis of \citet{1997Walsh}.
Figure \ref{fig3_4} presents the individual H{\sc{i}} channel maps of \mbox{NGC 1566} from ASKAP observations
in the velocity range between $1628$-$1393\,$km\,s$^{-1}$ and with a step-size  of $16\,$km\,s$^{-1}$.
Figure \ref{fig34c} shows the H{\sc{i}} column density map of \mbox{NGC 1566} as obtained from ASKAP observations (top panel),
the archival ATCA observations (middle panel), and a difference between the two maps (bottom panel). To produce the 
difference map, we have convolved the ASKAP and ATCA maps to the same beam angular resolution.\footnote{We make use of the Source Finding Application \citep[{{\sc{sofia}}},][]{2015Serra} to produce these maps. 
We apply a $4\sigma$ detection limit and a reliability of 95 per cent.} ASKAP observations are as sensitive as the ATCA observations to low surface brightness features surrounding the outer 
disc of \mbox{NGC 1566}. This figure emphasises the ability of the ASKAP instrument to probe relatively low column
density levels while mapping large areas of the sky. The small flux difference between
ASKAP and ATCA in the centre is likely due to the smaller baseline of the ASKAP configuration in comparison with that of the 
ATCA, however this difference is within the error of the two instruments.
Hence, ASKAP will provide the H{\sc{i}} community with unprecedented amount of high quality H{\sc{i}} line cubes.
We note that \citet{2016Reeves} presented the moment zero and velocity maps of \mbox{NGC 1566} using more recent ATCA observations
but with less integration time ($\sim17\,$hrs) in comparison with the archival ATCA  data used in this paper ($\sim36\,$hrs).
The study of \citet{2016Reeves} was mainly focused on the intervening H{\sc{i}} absorption in \mbox{NGC 1566} along with other nine nearby galaxies.\\

\begin{figure}
\begin{center}
\includegraphics[width=0.47\textwidth]{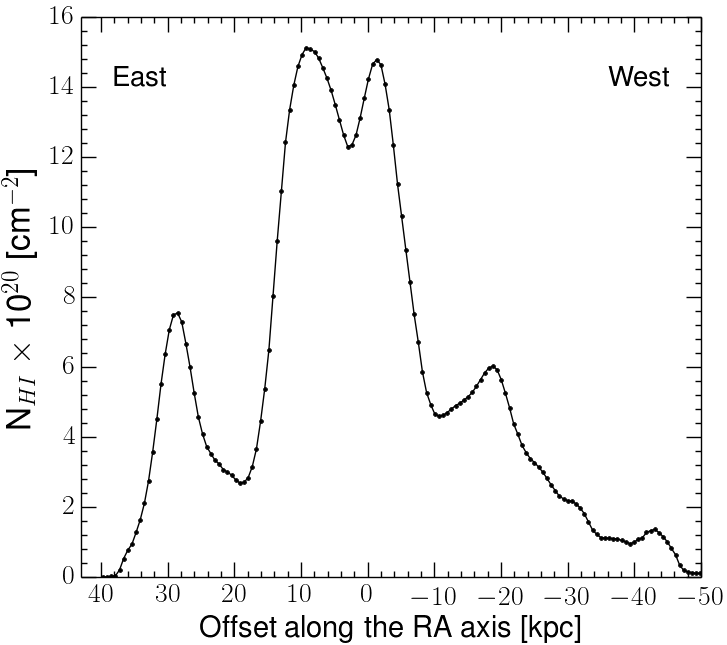}
\caption{A column density  cut across \mbox{NGC 1566} with respect to right ascension offset from 
the centre of the galaxy. The width of this cut is $5\,$kpc and is shown
by the blue-dashed line in Figure \ref{fig3_2}. Left-to-right in the figure corresponds to east-to-west.}
\label{fig3_3}
\end{center}
\end{figure}

\begin{figure*}
\begin{center}
\includegraphics[width=0.494\textwidth]{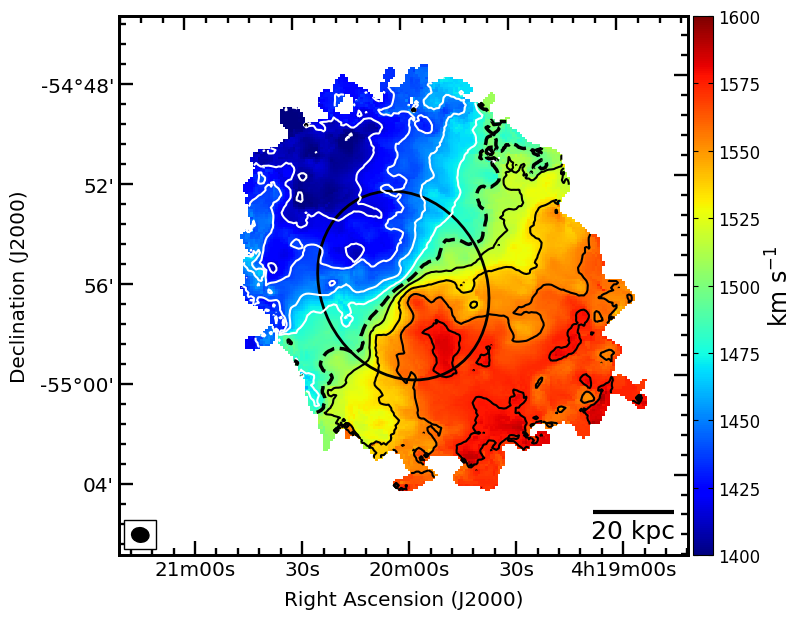}\includegraphics[width=0.479\textwidth]{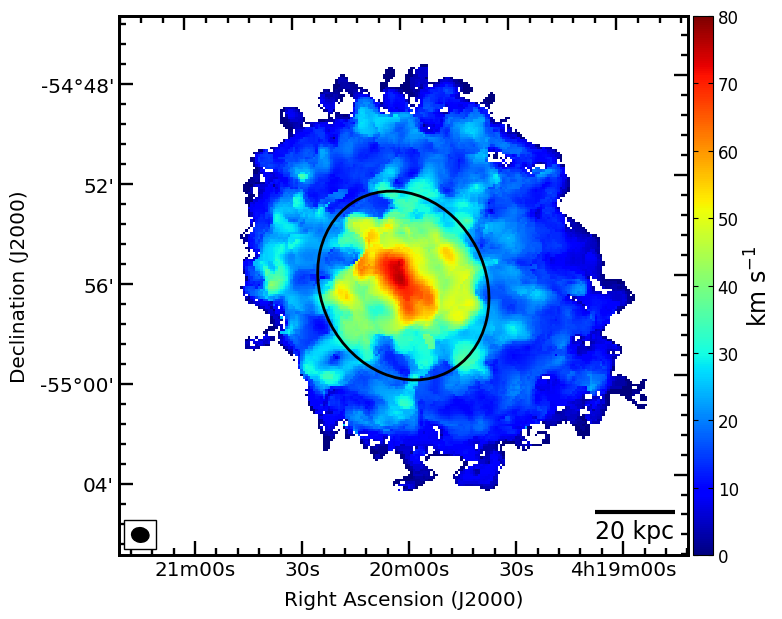}  \\
\caption{The ASKAP line-of-sight velocity (left) and velocity dispersion (right) of NGC 1566. The ellipse marks the optical 
disc of \mbox{NGC 1566} \citep{1997Walsh}. Contours are centred at the systemic velocity of NGC 1566 ($1496\,$km s$^{-1}$; black-dashed contour) and 
are equally separated by $20\,$km s$^{-1}$ interval; white and black contours show the approaching and receding sides, respectively. The black filled ellipse shows the restored beam of the observations.}
\label{fig3_4b}
\end{center}
\end{figure*}

Figure \ref{fig3_2} shows the optical DSS blue band image of \mbox{NGC 1566} with column density contours from ASKAP observations overlaid 
at $(0.6,1.2,2.4,4.8,9.6,14.0, 16.2)\times10^{20}\,$cm$^{-2}$. This map shows an H{\sc{i}} disc that extends beyond 
the observed optical disc especially around the northern and western parts of \mbox{NGC 1566} (also refer
to Figure \ref{fig5_deep}). The H{\sc{i}} gas is very concentrated in the inner arms and gradually decreases following the outer arms of the disc. The H{\sc{i}} also highlights the difference between 
the two outer arms better than in the optical (Figure \ref{Fig1_1}). The eastern outer arm forms a regular arc shape that extends between {\textit{PA}}$\,=50^\circ$ until where the {\textit{PA}}$\,=\,260^\circ$;
here the PA is estimated from the north extending eastwards to the receding side of the major axis. 
However, the western arm is significantly shorter extending from {\textit{PA}}$\,=\,210^\circ$ to {\textit{PA}}$\,=\,330^\circ$, and appears less regular or disturbed between {\textit{PA}}$\,=\,270^\circ$ to $330^\circ$. 
The $1.5\times10^{19}\,$cm$^{-2}$ column density contour extends for a diameter of almost $16'$, which at
the adopted distance of \mbox{NGC 1566} translates to a diameter of $\sim99\,$kpc.
Figure \ref{fig3_3} shows a column density cut across \mbox{NGC 1566} with respect to right ascension 
offset from the centre and measured at the declination of the centre of the galaxy. The
width of this cut is $5\,$kpc and is shown by the blue-dashed line in Figure \ref{fig3_2}. We use the \textsc{karma} 
visualisation tool \textsc{kvis} to generate this column density cut across \mbox{NGC 1566} \citep{1996Gooch}.
The column density of the H{\sc{i}} gas slowly drops with radius, mainly due to presence of the outer arms in \mbox{NGC 1566} at radius 
$r\,\sim20$kpc; the column density falls off at $r\,\sim20\,$kpc by only a factor of two from the peak value at $r\,\sim6\,$kpc.
Further, Figure \ref{fig3_3} shows an asymmetry in the distribution of the H{\sc{i}} gas in this galaxy; the
eastern part of the H{\sc{i}} disc sharply declines after $\sim30\,$kpc from the centre as opposed to the 
western part which extends beyond $\sim30\,$kpc and smoothly declines up to a radius of $\sim50\,$kpc.
This asymmetry is more evident in Figure \ref{fig3_2}, in which the H{\sc{i}} contours show crowding on the east side of the galaxy
and are spread out on the west side. The column density asymmetries present in \mbox{NGC 1566} could be signs of ram pressure
interaction, gas accretion and/or past flyby interaction(s) with the other members of \mbox{NGC 1566} galaxy group. 
We discuss the possibility of these interaction scenarios in Section 5. \\

\begin{figure}
\begin{center}
\includegraphics[width=0.47\textwidth]{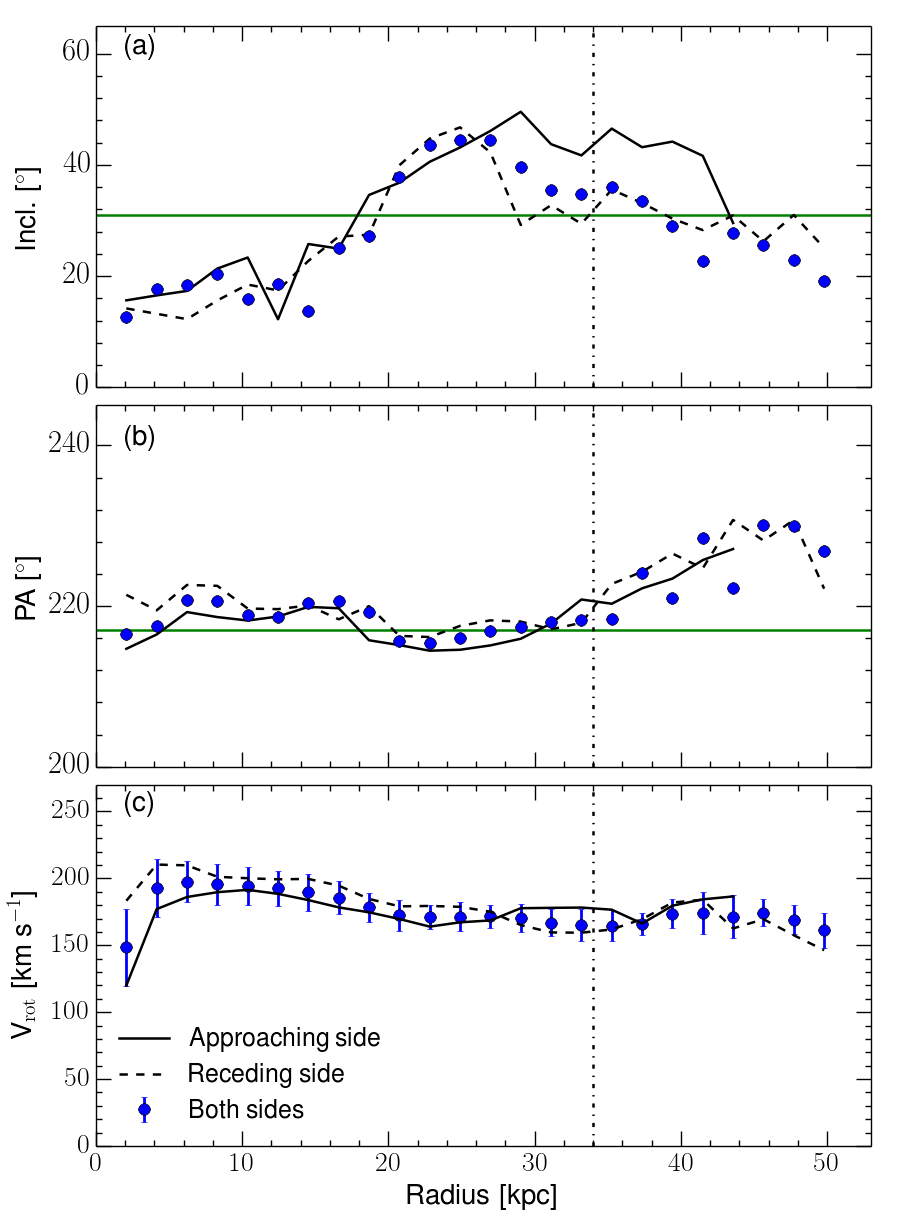}
\caption{The tilted ring model solutions for the inclination (a), the position angle (b), and the 
rotation velocity $V_{\x{rot}}$ (c) of \mbox{\mbox{NGC 1566}} using \textsc{rotcur}. 
Solid line, dashed line and filled blue circles denote the approaching, receding and both sides, respectively.
The green horizontal lines mark the best fit values for the {\textit{i}} ($\,31^\circ$) and {\textit{PA}}
($\,219^\circ$), respectively, which are used to estimate the rotation curve for both sides in the panel (c).
The vertical dash-dotted line shows the location of the optical radius of \mbox{NGC 1566} \citep{1997Walsh}.}
\label{fig_rot}
\end{center}
\end{figure}

\subsection{H{\sc{i}} KINEMATICS}
Previous studies of the kinematics of \mbox{NGC 1566} utilised optical spectroscopy to map the numerous emission line regions in this galaxy.
For example, \citet{1990Pence} used the Fabry-P\'erot interferometer at the $3.9\,$m Anglo-Australian Telescope to map the H$\alpha$ emission
in \mbox{NGC 1566} and were able to measure gas velocities out to a radius of $10\,$kpc \citep[c.f. Figures 4 \& 8 in][]{1990Pence}.
Figure \ref{fig3_4b} shows the ASKAP line-of-sight velocity and the velocity dispersion of \mbox{NGC 1566}.
The shape of the isovelocity contours of the approaching side (white contours) of the galaxy is slightly different in comparison with 
the receding side (black contours) which suggests the presence of kinematic asymmetry in NGC 1566.
The western side of the velocity field shows significant asymmetry in comparison with the eastern side.
This is also seen in both the moment zero map and the velocity channel map in Figures \ref{fig3_4} and \ref{fig34c}. In the individual channel maps,
the western side is more extended than the eastern side, especially at velocities between $1534-1456\,$km\,s$^{-1}$.
On the other hand, the dispersion map shows a very high peak in the inner regions which is likely a result of beam smearing.
There is a noticeable increase in the velocity dispersion associated with the inner arms of \mbox{NGC 1566}, similar to other grand design 
spirals like M83 \citep{2016Heald}. However the dispersion velocity decreases quickly in the outer arms and the outskirts of the HI disc.
We discuss in detail the possibility of an interaction scenario and other external influences that may lead to the lopsidedness in \mbox{NGC 1566}
in Section 5.\\

\begin{figure}
\begin{center}
\includegraphics[width=0.48\textwidth]{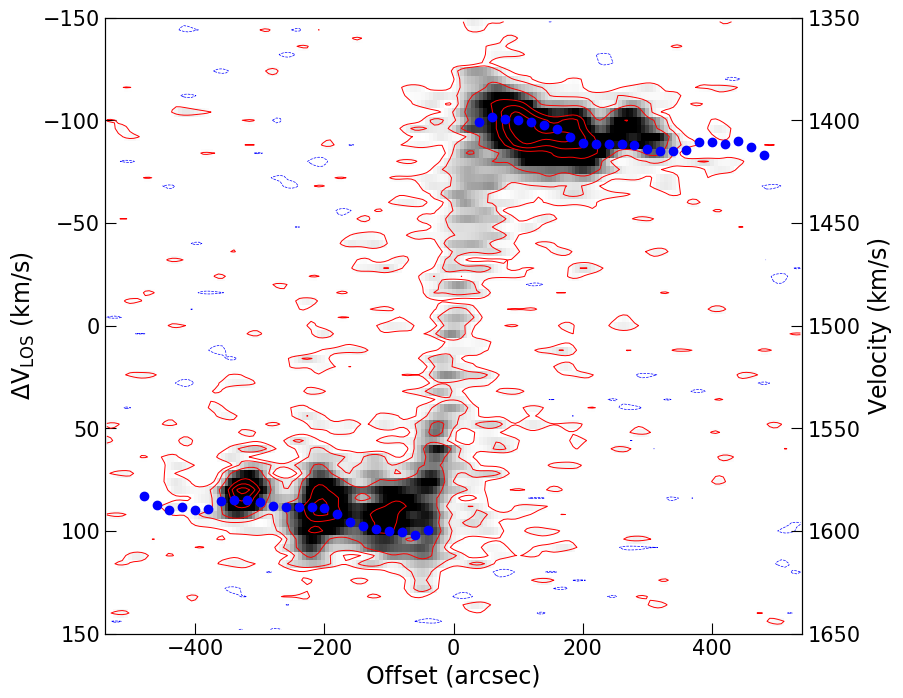}
\caption{The position-velocity diagram of \mbox{NGC 1566} along the major axis, overlayed is the rotation curve derived using {\sc{rotcur}}. 
The red contours are at $2, 4, 8, 16, 20\,$ times the $1\,\sigma$ noise level ($1.7\,$ mJy beam$^{-1}$ per $4\,$km\,s$^{-1}$ channel), 
while the blue contour is at $-2\,\sigma$.}
\label{fig3_3pv}
\end{center}
\end{figure}

\begin{figure}
\begin{center}
\includegraphics[width=0.47\textwidth]{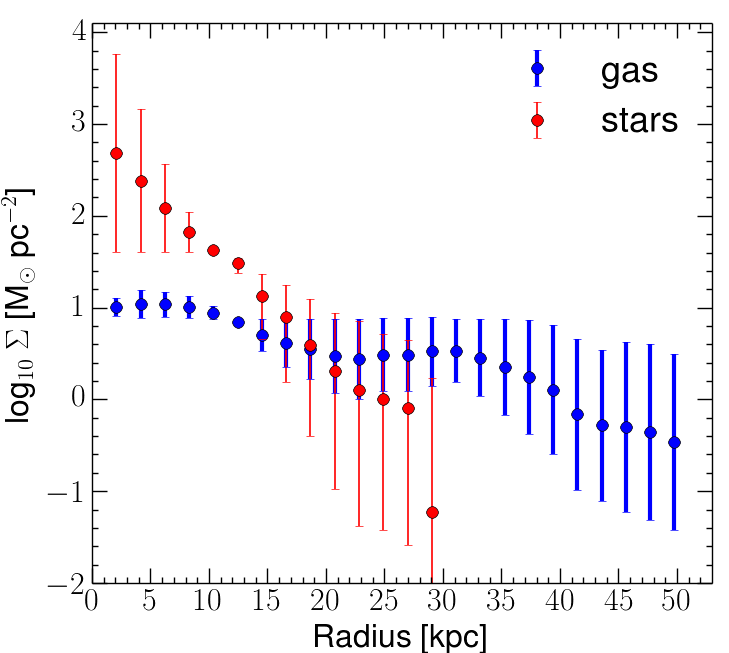}
\caption{The  gaseous and stellar radial mass surface density distributions
in \mbox{NGC 1566} shown in blue and red, respectively.}
\label{fig4_1}
\end{center}
\end{figure}

We follow the standard procedure described in \citet{2018Elagali}
to derive the H{\sc{i}} rotation curve of \mbox{NGC 1566} using a tilted ring model.
This model assumes that the gas moves in circular orbits. Each tilted ring is fitted independently 
as a function of radius and has $6$ defining kinematic parameters: the central
coordinate (x$_{c}$,y$_{c}$), the systemic velocity $V_{\x{sys}}$, the circular velocity $V_{\x{rot}}$, the inclination angle {\textit{i}},
as well as the position angle {{\textit{PA}}}. According to this model, the observed line-of-sight velocity V(x,y) is given by:

\begin{equation}
 V(x,y) = V_{\x{sys}} +  V_{\x{rot}}\, \sin i\, \x{cos}\,\theta,
\end{equation}

\noindent where the angle $\theta$ is a function of position angle and inclination. 
We use the Groningen Image Processing System \citep[\textsc{gipsy};][]{gipsy1,gipsy2} \textsc{rotcur} task \citep{rotcur1} to apply the
tilted ring model to the observed velocity fields of NGC 1566. We run the task in an iterative fashion to determine the above mentioned
free kinematic parameters for each ring, and use a ring width that equals half the restored beam-size ($\sim20$ arcsec), i.e., we fit two rings
per beam-size. Since the minor axis provides no information on the rotation curve, we apply $|\cos\,\theta|$ weighting function to minimise
the contribution of points far from the major axis. Firstly, we fit the systemic velocity $V_{\x{sys}}$ and the dynamical centre out to the edge of the optical disc simultaneously  
by fixing the {{\textit{PA}}} ($221^\circ$) and {\textit{i}} ($27^\circ$) to their optical values \citep{1990Pence}.
We next fit the inclination and position angle simultaneously, keeping the systemic velocity and the dynamical centre 
fixed to the determined values from the previous step. Then, we smooth the {{\textit{PA}}} and {\textit{i}} profiles with a radial boxcar 
function, estimate their average values and derive an optimum solution for the $V_{\x{rot}}$. The best fit for the systemic velocity  of \mbox{NGC 1566}
is $V_{\x{sys}}\,=\,1496\pm7\,$km\,s$^{-1}$, and the derived dynamical H{\sc{i}} centre agrees with the optical centre.
The position angle and inclination values estimated using \textsc{rotcur} are {{\textit{PA}}}$\,=\,219^\circ\pm4^\circ$ and
{\textit{i}}$\,=\,31^\circ\pm7^\circ$, respectively, and are within the error of the optical values determined by \citet{1990Pence}.
We derive the rotation curve for both the receding and approaching sides, to check for possible departures from symmetry and highlight any systematic uncertainties
associated with our final results. \\

Figure \ref{fig_rot} shows \textsc{rotcur} results for the inclination, the position angle and 
the rotation velocity of \mbox{NGC 1566}. The variation of the inclination with radius implies the existence of a mild warp in the H{\sc{i}} disc of this galaxy.
The inclination of the inner regions of the disc can be averaged to the value $19^{\circ}$ ($r =< 15\,$kpc), while the outer parts of
the disc ($r > 15\,$kpc) have an average inclination value of $37^{\circ}$. We refer the reader to Figure \ref{3dplot}, a 3D interactive visualisation of \mbox{NGC 1566}, 
the three axes in this visualisation are RA, Dec, and the velocity. On the other hand, the position angle derived using \textsc{rotcur} is constrained 
across the H{\sc{i}} disc. The green horizontal lines mark the best fit values for the {\textit{i}} ($\,31^\circ$) and {{\textit{PA}}
($\,219^\circ$), respectively, which are used to estimate the rotation curve for both sides in the lower panel.
The rotation of the approaching side is slightly different in comparison with the receding side's rotation.
Figure \ref{fig3_3pv} presents the position-velocity diagram of \mbox{NGC 1566} along the major axis,
with its rotation curve overlaid}. We note that the  low-level emission apparent in the forbidden quadrants
(gas with forbidden velocities) is not real and is due to sidelobes. The rotation of both sides  
reaches a velocity $V_{\x{rot}}\,=161\pm13\,$km\,s$^{-1}$ at a radius $r\,=\,50\,$kpc. This velocity translates to,
assuming a spherically-symmetric mass distribution, a total dynamical mass of $M_{tot}\,=\,(11.30\pm1.91$)$\times10^{11}\,M_{\odot}$
enclosed within this radius. As the fitted inclination of $31^{\circ}$ is outside the typical range where
\textsc{rotcur} is thought to be reliable \citep{rotcur1}, we also use the the Fully Automated 3D Tilted Ring Fitting Code \citep[\textsc{FAT};][]{2015Kamphuis} to derive the rotation curve of NGC 1566.  
This software works directly on the data cube, thus fitting in 3D, and is more robust against certain instrumental effects and hence 
is thought to be reliable to lower inclinations \citep{2015Kamphuis}. The results from \textsc{FAT} are consistent with those derived 
using \textsc{rotcur}. Hence, for brevity, we decide to only show and use the  \textsc{rotcur} results.\\

\section{Dark matter content and mass models}

The gravitational potential of any galaxy is a function of its combined gaseous, stellar and dark
matter mass components. In this section, we  investigate the distribution of the dark and baryonic matter 
in \mbox{NGC 1566}, and fit different mass models to the observed rotation curve of this system using the 
\textsc{gipsy} task \textsc{rotmas}. \textsc{rotmas} fits the following equation to the observed rotation ($V_{\x{obs}}$):

\begin{equation}
V_{\x{DM}}^2(r) = V_{\x{obs}}^2(r) - V_{*}^2(r) - V_{\x{gas}}^2(r),
\label{equation1}
\end{equation}
\noindent where $V_{\x{*}}$, $V_{\x{gas}}$, and $V_{\x{DM}}$ are the 
contributions of the stars, gas and the dark matter components to the total rotation curve of \mbox{NGC 1566},
respectively. The velocity contributions of the gaseous and the stellar mass components are estimated from 
their mass radial surface density distribution ($\Sigma({r})$)
using the \textsc{gipsy} task \textsc{rotmod} \citep{1983Casertano}. 
Below, we present the gaseous and the stellar mass radial surface density distributions along with
the different dark matter models used to fit the total rotation curve in equation \ref{equation1}.

\subsection{Gaseous Distribution}
To estimate the gaseous mass surface density profile, we use the H{\sc{i}} column density map obtained from our ASKAP observations (Figure \ref{fig3_2}). The gas surface density is measured in tilted rings using similar parameters
({\textit{i}}, {\textit{PA}}, dynamical centre) to those used to derive the total rotation curve with \textsc{rotcur}.
We use the  task \textsc{ellint} in \textsc{gipsy} to derive the radial H{\sc{i}} column density profile $N_{\x{H{\sc{i}}}}$($r$). 
We then convert the H{\sc{i}} column density profile to gaseous mass surface density. We scale the gas mass surface density by a factor of $1.4$
to account for the presence of helium and metals and assume that it is optically thin. Figure \ref{fig4_1} shows the 
gas mass surface density profile of \mbox{NGC 1566}. We use the radial gas mass surface density profile of \mbox{NGC 1566} to estimate the corresponding gas 
rotation velocities and assume that  the gas is mainly distributed \mbox{in a thin disc.}\\

\subsection{Stellar Distribution}
To derive the stellar mass surface density profile in \mbox{NGC 1566}, we use the infrared (IR) photometry obtained from 
the Infrared Array Camera (IRAC) on board the Spitzer Space Telescope \citep{2004Werner, 2004Fazio}. 
We convert the IR radial flux density distribution ($S_{\lambda}({r})$) to stellar radial mass density distribution
($\Sigma_{*}({r})$) using the following equation:

\begin{equation}
 \Sigma_{*}({r}) \sim S_{\lambda}({r}) Y_{\lambda},
\end{equation}

\noindent where $\lambda$ indicates the wavelength band and $Y_{\lambda}$ is the mass-to-light ratio with respect to that band. Here, we derive the surface density 
profile in two different bands, namely, the IRAC $3.6$ and $4.5\,\mu$m bands. We follow the approach in \citet{2008seheon} to derive the mass-to-light ratio of the IRAC
two near-infrared bands; for \mbox{NGC 1566} the values are $Y\,$=$\,0.648\pm0.073$ and $0.611\pm0.072$ for the $3.6$ and $4.5\,\mu$m bands, respectively.
Similar to the gaseous profile, we use \textsc{ellint} to measure the near-infrared flux density in tilted rings  with the same parameters ({\textit{i}}, 
{\textit{PA}}, dynamical centre) used to derive the rotation curve of NGC 1566. \\

\noindent We then convert the near-infrared flux density from the IRAC pipeline flux units (MJy sr$^{-1}$) to solar units and apply aperture correction 
for the $3.6$ and $4.5\,\mu$m flux density. The stellar mass radial surface density $\Sigma_{*}({r})$, is calculated by the following equation:

\begin{equation}
 \Sigma_{*}({r}) = C_{\lambda} Y_{\lambda} f_{\lambda} \frac{S_{\lambda}({r})} {ZP},
\end{equation}

\noindent where $f_{\lambda}$ is the aperture correction factor, $C_{\lambda}$ is the conversion factor and $ZP$ is the zero point magnitude for each band. We 
adopt aperture correction values of $f_{3.6\mu\x{m}}\,=\,0.944$ and $f_{4.5\mu\x{m}}\,=\,0.937$ and zero point magnitude fluxes of $280.9$ and $179.7\,$Jy for the $3.6$ and $4.5\,\mu$m 
\citep{2005Reach}, respectively. Following the calculations in \citet{2008seheon} based on the spectral-energy distributions of the Sun, we use $C_{3.6\mu\x{m}}\,=\,0.196$ and
$C_{4.5\mu\x{m}}\,=\,0.201\,$M$_{\odot}$pc$^{-2}$. Figure \ref{fig4_1} shows the stellar mass radial surface density profile of \mbox{NGC 1566}. 
We measure a total stellar mass of ($6.5\pm0.4$)$\,\times10^{10}\,$M$_{\odot}$, using the averaged $3.6$ and $4.5\,\mu$m bands 
stellar mass radial surface density profiles, which is within the error of the value reported in \citet{2014Laine}. The stellar mass radial profile can be described by a simple exponential function ($\Sigma_{*}\sim \x{exp}(-r/h)$), for which the the radial scale-length ($h$)
equals $\,3\,$kpc and thus the scale-height of the disc ($z_{0}$), using $h/z_{0}\,\simeq\,5\,$\citep{2002Kregel,1981Kruit}, equals $0.6\,$kpc.
Using \textsc{rotmod}, we construct the stellar velocity component from the stellar mass radial surface density profile of \mbox{NGC 1566}, assuming that the stellar disc has a vertical
sech$^2$ scale-height distribution with $z_{0}\,=\,0.6\,$kpc \citep{1981Kruit}. 

\subsection{Dark Matter Halo Profiles}
\subsubsection{Pseudo-isothermal Dark matter Profile}
This profile is the simplest and most commonly used in the studies of
galaxies' rotation curves \citep{1987Kent,1991Begeman}. 
This model assumes a central constant-density core ($\rho_0$ ) and a density profile given by:
\begin{equation}
 \rho({r}) = \frac{\rho_0}{1+(r/r_c)^2},
 \end{equation}

\noindent where  $r_c$ is the core radius. The corresponding rotation velocity ($v(r)$) to
the pseudo-isothermal (ISO) potential is:

\begin{equation}
 v^2({r}) = 4\pi G \rho_0 r_c^2 \Big[1-\frac{r_c}{r} \arctan\Big(\frac{r}{r_c}\Big)\Big],
 \end{equation}
\citep{1986Kent}.
\subsubsection{Burkert Dark matter Profile}
This profile adopts the following definition for the dark matter density profile \citet{1995Burkert}:
\begin{equation}
 \rho({r}) = \frac{\rho_0 r_c^3}{(r+r_c)(r^2+r_c^2)},
 \end{equation}
\noindent Similar to the ISO profile, $r_c$ and $\rho_0$ denote the core radius and the central density of 
the dark matter halo, respectively. This profile resembles the distribution expected for a pseudo-isothermal sphere at the inner radii ($r << r_c$) and predicts central density $\rho_0$. At larger radii,
the Burkert density profile $\rho({r})$ is roughly proportional to $r^{-3}$. The velocity corresponding to this profile is given by 
\citep{2000Salucci}: 

\begin{equation}
 v^2({r}) = \frac{6.4G\rho_0 r_c^3}{r}\Big[\ln\Big(1+\frac{r}{r_c}\Big)+\frac{1}{2}\ln\Big(1+\frac{r^2}{r_c^2}\Big)-\arctan\Big(\frac{r}{r_c}\Big) \Big].
\end{equation}
 
\noindent To fit the rotation curve of this halo profile, we have two free parameters, namely, the 
core radius and the central density of the dark matter halo.

\begin{figure}
\begin{center}
\includegraphics[width=0.45\textwidth]{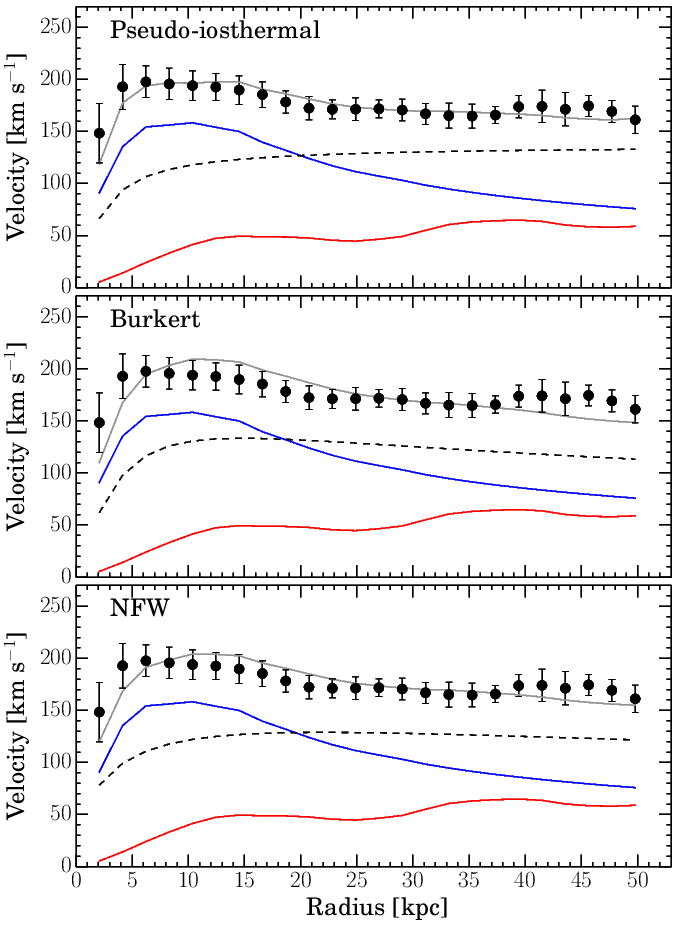}
\caption{The mass models fit to the observed rotation curve of \mbox{NGC 1566} using the ISO,
Burkert and NFW dark matter profiles. The black circles and  grey line show the observed rotation curve and the 
best fit modelled rotation curve of NGC 1566, respectively. The red, blue and dashed-black lines show the contributions
of the gaseous, stellar and dark matter components to the overall rotation curve of NGC 1566, respectively.}
\label{fig4_2}
\end{center}
\end{figure}
\subsubsection{NFW Dark matter Profile}
\citet{1996Navarro, 1997Navarro} used N-body simulations to explore the equilibrium density profile of the dark matter
halos in a hierarchically clustering universe and found that these profiles  have  the same shape, regardless
of the values of the cosmological parameters or the initial density fluctuation spectrum. The density profile in this case can be
described by the following equation:

\begin{equation}
 \rho({r}) = \frac{\delta_c \rho_{crit}}{\frac{r}{r_s}\Big(1+\frac{r}{r_s}\Big)^2},
\end{equation}

\noindent where $r_\x{s}$ is the  scale radius, $\delta_c$ is a critical dimensionless density of the halo and $\rho_{\x{crit}}$
is the critical density for closure. The NFW halo profile is similar to the Burkert profile; 
the only difference is at $r << r_s$, in which the NFW halo density $\rho({r})\sim r^{-1}$ instead of a constant
core density value as it is the case for the Burkert profile. The velocity of this profile is given by:
\begin{equation}
 v^2({r}) = \frac{v_{200}^2}{x}\frac{\ln(1+cx)-\frac{cx}{1+cx}}{\ln(1+c)-\frac{c}{1+c}},
\end{equation}
 
\noindent where $x$ is the radius ($x\,=\,r$/$r_{200}$) in virial radius units, $c$ is the halo concentration $c\,=\,r_{200}$/$r_s$ and $v_{200}$ is the 
circular velocity at $r_{200}$:

\begin{equation}
 v_{200} = \sqrt{\frac{G M_{200}}{r_{200}}}\,=\,10 H_0 r_{200}.
\end{equation}

\noindent Here, $H_0$ is the Hubble constant and $M_{200}$ is the virial mass. To fit the rotation curve of the NFW halo 
profile, we have two free parameters, namely, the 
scale radius and the virial radius of the dark matter halo.

\subsection{Mass Model Results}
To derive the dark matter contribution to the total rotation curve of \mbox{NGC 1566}, we fit equation \ref{equation1}
using \textsc{rotmas}. Figure \ref{fig4_2} shows the mass model results
using the ISO, Burkert and the NFW halo profiles. Table 6 lists the results of our mass models. 
In all models, the stellar disc dominates the rotation curve up to a radius $r\sim18\,$kpc and starts to sharply 
decline at $r \gtrsim 20\,$kpc. On the other hand the dark matter rises linearly with radius reaching the maximum at $r\sim10\,$kpc and remains 
fairly constant at larger radii. Based on the ISO, Burkert and the NFW halo profiles, we estimate dark matter fractions in \mbox{NGC 1566} to be $\simeq0.66$, $\simeq0.58$ 
and $\simeq0.62$, respectively. The three  dark matter profiles result in reasonable fits. However, due to the lack of angular resolution in the inner regions of \mbox{NGC 1566} ($\sim2\,$kpc),
we can not differentiate between the three dark matter density profiles. To distinguish between different dark matter halo profiles, higher angular resolution observations
of few hundred parsec scales are required \citep[see for example][]{2015seheon,2010deBlok, 2002Bolatto, 2001deBlok}.
This question will soon be addressable for large samples of nearby galaxies using the $36$ ASKAP antennas 
(longest baseline of $6\,$km) and MeerKAT telescope \citep{2016deblok}, which will aid constraining the mass distributions
of both the dark and baryonic matter in large numbers of galaxies.\\

\begin{table*}
\begin{center}
\label{table2}
\begin{tabular}{llll}
\multicolumn{2}{l}{{\bf{Table 6.}} Mass models fit results for \mbox{NGC 1566}.} \\ \hline \hline
Parameter& Pseudo-isothermal & Burkert &NFW \\ \hline 
$r_\x{c}$ (kpc) & $1.9\pm0.6$&$4.3\pm0.8$  & - \\
$\rho_\x{0}$ (M$_{\odot}$ pc$^{-3}$)&$0.092\pm0.050$ &$0.082\pm0.037$ & - \\
$r_\x{s}$ (kpc) &- & - & $9.52\pm1.78$\\
$r_{\x{200}}$ (kpc)& -& - & $80.84\pm3.71$\\
M$_{\x{DM}}$ ($10^{11}$M$_{\odot}$)&$7.60$ &$5.38$ &$6.29$ \\
$\chi_{\x{red}}^2$ & $0.34$& $1.36$& $0.65$\\
$f_{\x{DM}}$ &$0.66$ & $0.58$  & $0.62$\\ \hline
\end{tabular}
\end{center}
\end{table*}

\subsection{Tully-Fisher Distance of NGC 1566}
Here, we use the Tully-Fisher (TF) relation \citep{1977Tully} in an attempt to measure a 
more accurate distance for NGC 1566 than currently reported in the literature. 
We use the $I$-band apparent magnitude ($\x{m}_{I}$) and line width ($w_{50}$) values reported in Table 1 \& 3, respectively,
and the kinematic inclination derived in Section 3.2. The distance of NGC 1566 using the $I$-band TF relation \citep{2006Masters} is $16.9^{+7.5}_{-4.1}\,$Mpc, which is smaller than 
the value we adopt but within the errors. Unfortunately, the error in the TF distance does not allow us to definitively exclude the far distance value reported in NED
nor claim a superior distance measurement for this galaxy. We have therefore left the nominal distance of this galaxy
as $21.3\,$ Mpc throughout this paper. The large error in the TF relation is due to the low inclination of \mbox{NGC 1566}, which is 
also reflected by the large uncertainties in the distance measurements for this galaxy in the literature.

\section{Discussion}
\subsection{Possible Origins for the H{\sc{i}} Disc Asymmetries}
Many disc galaxies are asymmetric, and have lopsided stellar and/or  gaseous  
components \citep{1980Baldwin,2005Bournaud,2008Mapelli}.
Both theoretical and observational studies suggest 
three  different environmental mechanisms  that can  cause such asymmetries, namely, ram pressure interactions with 
the IGM, galactic  interactions as well as gas  accretion from hosting/neighbouring filaments 
 \citep{2007Oosterloo, 2007McConnachie,2008Reichard,2008Mapelli,2014Yozin,2014deBlok,2018Vulcani}.
In this subsection, we explore the possibility of each of these scenarios given the available data for \mbox{NGC 1566}. 
Below, we combine the discussion of the galactic interactions and gas accretion scenarios in one subsection for brevity while discussing 
the ram pressure stripping scenario separately.

\subsubsection{Interactions and/or Accretion Scenarios}
\begin{figure*}
\begin{center}
\includegraphics[width=0.87\textwidth]{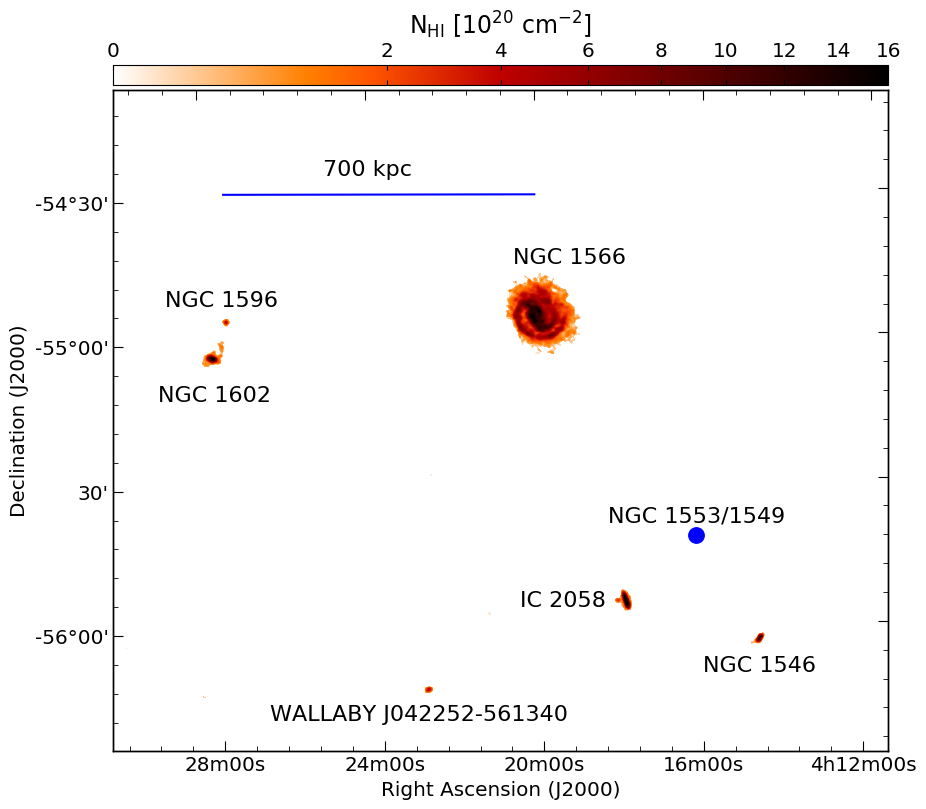}
\caption{The H{\sc{i}} column density map of \mbox{NGC 1566} mosaic field as revealed by ASKAP early science observations.
The synthesised beam  size is $\theta_{\x{FWHM}}\,=\,42''\times35''$. There are six galaxies detected in this mosaic, within a velocity range between
$950-2040$ km s$^{-1}$. The projected physical distance between \mbox{NGC 1566} and the galaxy pair NGC 1596/1602 is shown by 
the blue horizontal line. The blue filled circle shows the centre of the group, and the location of the two interacting galaxies NGC 1553/1549 \citep{2005Kilborn,2017ATully}.}
\label{fig5_map}
\end{center}
\end{figure*}

Tidal interactions between \mbox{NGC 1566} and neighbouring galaxies could lead to its asymmetries and warps. 
The interacting galaxy pair NGC 1596/1602 \citep{2006Chung} is a possible candidate for such a tidal 
encounter, the physical projected separation between this galaxy pair and \mbox{NGC 1566} is $\sim700\,$kpc ($2^{\circ}$ on the sky).
Figure \ref{fig5_map} shows the column density map of \mbox{NGC 1566} mosaic field. We detect six galaxies including \mbox{NGC 1566} in this mosaic.
A detailed description of these galaxies and the remainder of the Dorado group galaxies will be presented in 
Elagali et al. (in prep.) and Rhee et al. (in prep.). 
Even though  tidal interactions between \mbox{NGC 1566} and the neighbouring galaxies may have occurred,
we do not see any H{\sc{i}} tail/bridge that would result from such an interaction. The centre of the \mbox{NGC 1566} group is shown by the 
blue filled circle in Figure \ref{fig5_map}, as defined by where the two most massive and bright galaxies of this group are located, namely
NGC 1553/1549 \citep{1984Kormendy,2005Kilborn,2017ATully}.
We detect no H{\sc{i}} emission above the rms noise of ASKAP observations from this interacting galaxy pair. 
The upper H{\sc{i}} mass limit of NGC 1553/1549 based on  the $3\sigma$  noise level of this observations 
and over $40\,$km s$^{-1}$ channel widths (ten channels) is $\simeq 3.7\times10^{6}\,$M$_{\odot}$ beam$^{-1}$ at
the distance of this galaxy pair \citep[$17.6\,$Mpc,][]{2005Kilborn}.\\

\begin{figure}
\begin{center}
\includegraphics[width=0.48\textwidth]{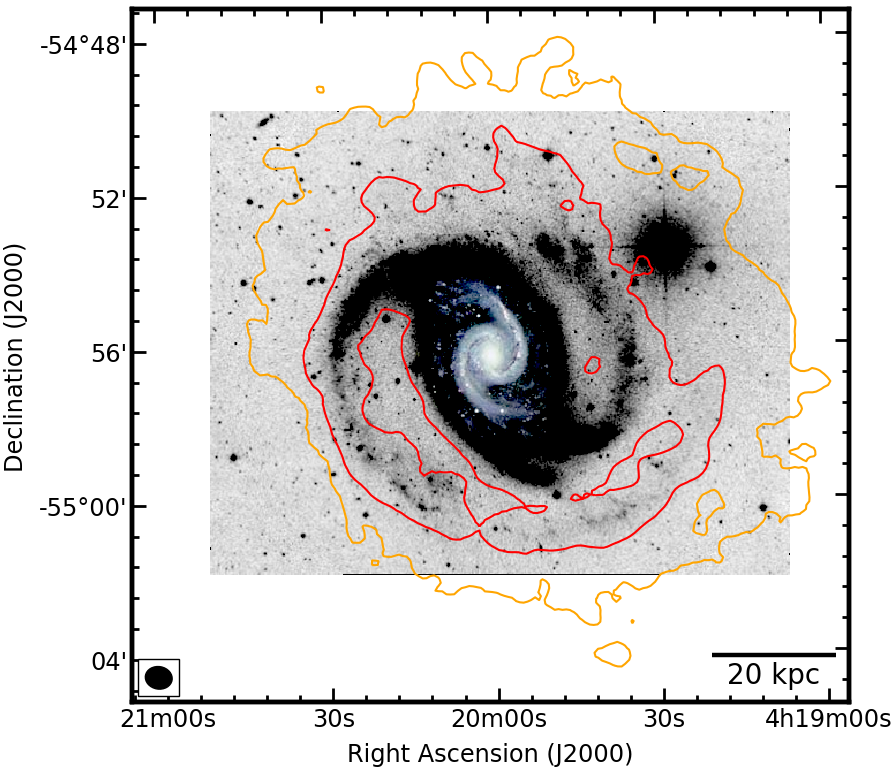}
\caption{Deep optical image of \mbox{NGC 1566} observed by David Malin and available at the Australian Astronomical Observatory database. 
The red and orange contours are at  H{\sc{i}}  column density values of $N_{\x{H{\sc{i}}}}\,=\,3.7\times10^{20}\,$cm$^{-2}$ 
and $\,0.6\times10^{20}\,$cm$^{-2}$, respectively. The black ellipse shows the restored beam of the ASKAP observations.}
\label{fig5_deep}
\end{center}
\end{figure}

Figure \ref{fig5_deep} presents a deep optical image of \mbox{NGC 1566} taken by David Malin at the  
Australian Astronomical Observatory. We also examine this deep image and see no faint stellar substructures 
around \mbox{NGC 1566}, nor a dramatic system of streams/plumes that may have formed through a tidal interaction or minor merger with neighbouring galaxies; see 
for example \citet{2018Kado,2018Pop,2018Elagali,2015Mart,2014deBlok2ww, 2010Mart}. We note that these faint structures  can  form  on timescales of $10^{8}$ years and  exist for few gigayears after
the interaction \citep{1988Hernquist2,1989Hernquist1}. Hence, a recent interaction scenario is less likely to be the reason for the asymmetries
seen in the outer H{\sc{i}} disc of \mbox{NGC 1566}. Alternatively, the lopsidedness of the H{\sc{i}} distribution  of \mbox{NGC 1566} can be a result 
of gas accretion. \citet{2014deBlok} present an example of this scenario, in which a low column density, extended cloud  is connected to the
observed  main H{\sc{i}} disc of NGC 2403. In the case of \mbox{NGC 1566}, we do not detect any clouds or filaments connected to, or in the 
nearby vicinity of, the H{\sc{i}} disc of this galaxy. We note that our $3\sigma$ column density sensitivity measured over a $20\,$km s$^{-1}$ width
is $N_{\x{H{\sc{i}}}}\,=\,3.7\times10^{19}\,$cm$^{-2}$. Therefore, it is difficult to say any more quantitative/conclusive statements about the accretion effects on the H{\sc{i}} disc of \mbox{NGC 1566}
and that more sensitive H{\sc{i}} observations of this galaxy are needed to rule out an accretion scenario.\\

\subsubsection{Ram Pressure Stripping Scenario}
Ram pressure is widely observed in massive galaxy clusters \citep{1991White, 1999Abadi, 2003Acreman, 2008Randall, 2009Vollmer, 2016Merluzzi, 2017Ruggiero, 2017Sheen}, 
such as the  nearby Virgo cluster \citep{2007Chung, 2017Yoon}, and is the main reason for the stripping and removal of gas in  galaxy
clusters especially closer to their dynamical centres. Even though ram pressure stripping 
is not as prevalent in galaxy groups, a few cases have been reported in the literature 
\citep{Kantharia2005, 2007McConnachie, Westmeier-2011, 2012Rasmussen, 2016Heald,2018Vulcani}.
Many authors also ascribe ram pressure as a potential cause for H{\sc{i}} deficiency in groups \citep[see for example][]{Sengupta2006,2007Sengupta,2010Freeland,2016denes}.
Here, we investigate the asymmetries present in the outskirts of  \mbox{NGC 1566} and its
connection to the gaseous halo undergoing ram pressure stripping as a consequence of its interaction with the IGM.
As a galaxy passes through the IGM with an inclined orientation, the  H{\sc{i}}  gas will be compressed in the leading 
edge of the outer disc while the gas at the lagging edge get stripped and pulled away as a result of the ram-pressure forces \citep[refer to Figure 1 and 3 in][]{2000Quilis}.
This will produce an asymmetry in the H{\sc{i}} column density distribution similar to that observed
in \mbox{NGC 1566}, in which the south-eastern edge of the H{\sc{i}} disc sharply declines after $\sim30\,$kpc from the centre, 
while the north-western edge is more extended 
and smoothly declines with radius up to $\sim50\,$kpc (refer to Figures \ref{fig3_2} \& \ref{fig3_3}).\\

\begin{figure}
\begin{center}
\includegraphics[width=0.46\textwidth]{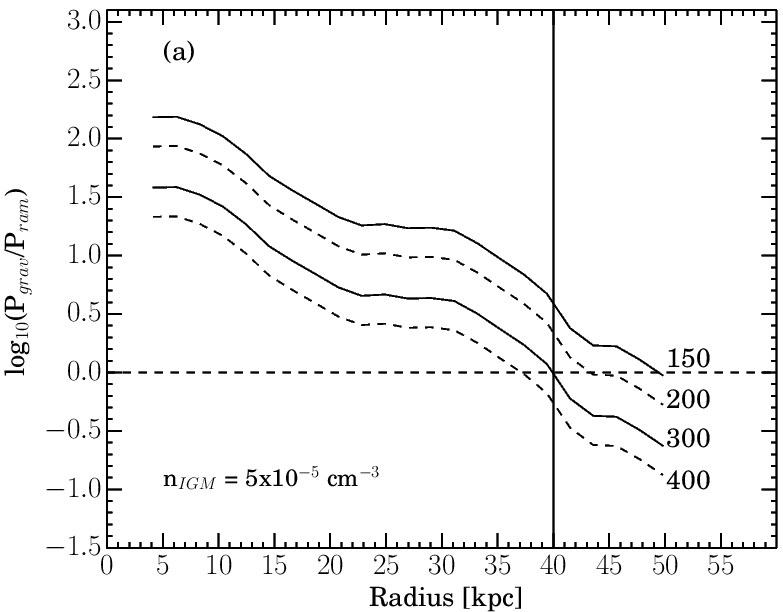}\\
\includegraphics[width=0.46\textwidth]{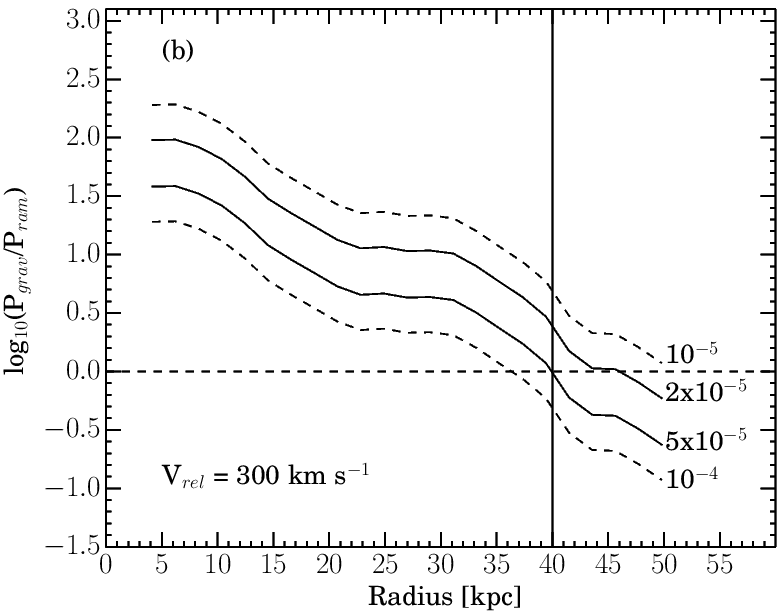}  
\caption{A comparison between the gravitational pressure and ram pressure ratio in the disc of \mbox{NGC 1566}
for different radii. (a) the pressure ratio assuming an IGM density value of n$_{\x{IGM}}\,=\,5\times10^{-5}\,$cm$^{-3}$ 
and relative velocities for \mbox{NGC 1566} through the IGM of v$_{\x{rel}}\,=\,150,\,200,\,300$ and $400\,$km s$^{-1}$.
(b) the pressure ratio for  constant  velocity  (v$_{\x{rel}}\,=\,300\,$km s$^{-1}$) and four 
different IGM density values (n$_{\x{IGM}}\,=\,1.0\times10^{-5},\,2\times10^{-5},\,5\times10^{-5}\,$ and
$1.0\times10^{-4}$cm$^{-3}$). The horizontal dashed line represents the equality, above which the gaseous disc is dominated by 
gravity and below  where ram pressure dominates and the disc is subjected to ram pressure stripping. 
The vertical line shows the radius at which the H{\sc{i}} column density of \mbox{NGC 1566} is lopsided; 
At r$\,=\,40\,$kpc the column density of the south-eastern part of the H{\sc{i}} disc is nearly zero, while the north-western edge is more extended and smoothly
declines with radius up to $\sim50\,$kpc.}
\label{fig3_4c}
\end{center}
\end{figure}

To examine the link between the observed asymmetries in the gas disc of \mbox{NGC 1566} and ram pressure stripping, we follow the approach proposed by \citet{1972GunnGott}
and compare the restoring force by the gravitational potential at the outer disc and the pressure from the IGM.
The gas in the outer parts will remain intact to the halo as long as the ram pressure force is lower in
magnitude than the pressure from the gravitational potential of the halo. 
For a disc galaxy with a gravitational potential $\phi(r)$ at a distance $r$ from the centre, the restoring force due to this gravitational 
potential will exert a pressure given by the following equation \citep{2018Kppen,2006Roediger,2005Roediger}:

\begin{equation}
 P_{\x{grav}} = \Sigma_{\x{gas}}({r}) \Big|\frac{\partial \Phi(r)}{\partial z}\Big|_{\x{max}},
\end{equation}

\noindent where the derivative denotes the maximum value of the restoring force at height $z$ above the disc. 
The height that corresponds to the maximum force is estimated by equating the second derivative of the gravitational potential
$\frac{\partial^2 \Phi(r)}{\partial z^2}$, to zero. On the other hand, the IGM will apply a pressure on the outer-most part of the disc that
is given by the formula:

\begin{equation}
 P_{\x{ram}} = \rho_{\x{IGM}} v_{\x{rel}}^2,
\end{equation}

\noindent where $\rho_{\x{IGM}}$ is the density of the IGM and $v_{\x{rel}}$ is the relative velocity of the galaxy. The ram pressure
is capable of removing gas from the galaxy as it passes through the IGM only when $P_{\x{ram}}$ > $P_{\x{grav}}$. For our calculation, we make 
two assumptions. First, we only calculate the gravitational force due to the dark matter and neglect the restoring force from the 
stars and gas, i.e., we assume that the contribution from the baryonic matter is negligible. This is only true at the  outer radii 
(refer to Figure \ref{fig4_2}) where the contribution of the dark matter dominates the total halo mass of \mbox{NGC 1566} 
\citep{Westmeier-2011}. The second necessary assumption is that \mbox{NGC 1566} moves with an 
intermediate or ``face-on'' vector into the IGM, when ram pressure stripping is efficient 
\citep{2000Quilis, 2001AVollmer,2005Roediger} but not directly in the line-of-sight, where morphological effects would
be harder to discern \footnote{We remind the reader that the inclination of \mbox{NGC 1566} 
with respect to the observer is known but the orientation of the disc relative to the direction of motion through the
IGM is unknown.}. Based on the HI morphology and kinematics of this galaxy an encounter with the IGM at an intermediate inclination is favoured. 
To estimate the restoring force acting on the disc of \mbox{NGC 1566}, we use the NFW dark matter halo gravitational potential which is described by the equation: 

\begin{equation}
\phi_{\x{NFW}}(r)=-\frac{G M_{\x{200}} \x{ln}\Big(1+\frac{r}{r_s}\Big)}{r\Big[\x{ln}(1+c)-\frac{c}{1+c}\Big]}.\\
\end{equation}

Figure \ref{fig3_4c} presents the results of our simple analytic model, in which the ratio between $P_{\x{grav}}$ and $P_{\x{ram}}$ is calculated
at different radii.  In Figure \ref{fig3_4c}a, we calculate this ratio assuming that the IGM density is constant (n$_{\x{IGM}}\,=\,5\times10^{-5}\,$cm$^{-3}$) and 
use different relative velocities for \mbox{NGC 1566}. This IGM density is similar to the Local Group gas density value \citep{2001Rasmussen, 2005Williams}.
Hence, we note that this adopted IGM density is a lower limit to the IGM density of \mbox{NGC 1566} galaxy group. The \mbox{NGC 1566} group has a 
halo mass  \citep[M$_{\x{halo}}\sim10^{13.5}\,$M$_{\odot}$;][]{2005Kilborn} larger than the Local Group and consequently hotter/denser IGM gas is expected
\citep{1998Eke,2009Pratt,2017Barnes}. In Figure \ref{fig3_4c}b, we derive the same ratio adopting a constant relative velocity for \mbox{NGC 1566} (v$_{\x{rel}}\,=\,300\,$km s$^{-1}$) 
and different values for the IGM gas density. We use N-body simulations to predict the probability distribution of the relative velocity of \mbox{NGC 1566} in the IGM following the 
orbital libraries described in \citet{2013Oman, 2016Oman} and based on the projected coordinates of \mbox{NGC 1566} (angular and velocity offsets)
from the group centre \citep[refer to][for more information on this group]{2005Kilborn}. The relative velocity of a subhalo (galaxy) with a 
similar mass to \mbox{NGC 1566} falling into a host halo (galaxy group), with a similar mass to \mbox{NGC 1566} group, based on this analysis lies 
in the range between $250-500\,$km s$^{-1}$ at $99\,$per cent confidence. Figure \ref{fig3_4c} shows that the outer part of the H{\sc{i}} disc of \mbox{NGC 1566} ($r\gtrsim40\,$kpc), for certain values for v$_{\x{rel}}\,$ and n$_{\x{IGM}}$, can be affected by ram pressure 
winds in particular for n$_{\x{IGM}}\gtrsim5\times10^{-5}\,$cm$^{-3}$ and v$_{\x{rel}}\gtrsim200\,$km s$^{-1}$. 
The highest IGM density adopted for NGC 1566 group and used in Figure \ref{fig3_4c}b (n$_{\x{IGM}}\gtrsim10^{-4}\,$cm$^{-3}$)
is consistent and in lower bound of the IGM density values reported for loose galaxy groups and derived from x-ray luminosities in 
\citet{Sengupta2006,2010Freeland}.
As expected for the inner regions the pressure due to the gravitational potential is much higher than the ram pressure force. 
We note that the ratio in the inner radii is  higher than shown in the figure since, as already noted, 
we  neglect the contributions from the stellar and the gaseous gravitational potentials (refer to Figures \ref{fig4_1}-\ref{fig4_2}).\\

Even though our simple analytic approach suggests that ram pressure interaction with the IGM is the likely reason for the 
lopsidedness of the gas morphology of \mbox{NGC 1566}, the result is tentative.
This is for two reasons. First, our ASKAP early science observation is not sensitive enough to probe H{\sc{i}} column densities below $10^{18}\,$cm$^{-2}$, which 
means that we can not rule out gas accretion as a plausible reason for the asymmetries seen in \mbox{NGC 1566}. 
The second caveat is that all the environmental processes that affect the H{\sc{i}} gas in galaxy groups such as
tidal stripping by the host-halo (group), galaxy-galaxy encounters and ram pressure interaction operate at similar radial distances from the group centre. This is to say, as a galaxy approaches the centre of the host-halo and is within 
a distance of d/R$_{\x{vir}}$ < $0.5$, where d is the physical distance from the group centre and R$_{\x{vir}}$ is the virial radius of the group, it can experience ram pressure stripping from the IGM and/or 
can also tidally interact with nearby satellites \citep{2016Marasco,2013yannick}. This adds to the complexity of disentangling the contributions
of different external processes on the H{\sc{i}} gas content and morphology in galaxy groups. However, we think that 
sensitive H{\sc{i}} observations of large samples of galaxies,
similar to those that WALLABY, APERTIF and MeerKAT will deliver in the next few years, 
will provide the H{\sc{i}} community with the opportunity to systematically study the H{\sc{i}} gas in group environments and help disentangling
the contributions of these environmental processes.\\


\subsection{Gas Content and Star Formation Rate in \mbox{NGC 1566}}
\begin{figure}
\begin{center}
\includegraphics[width=0.47\textwidth]{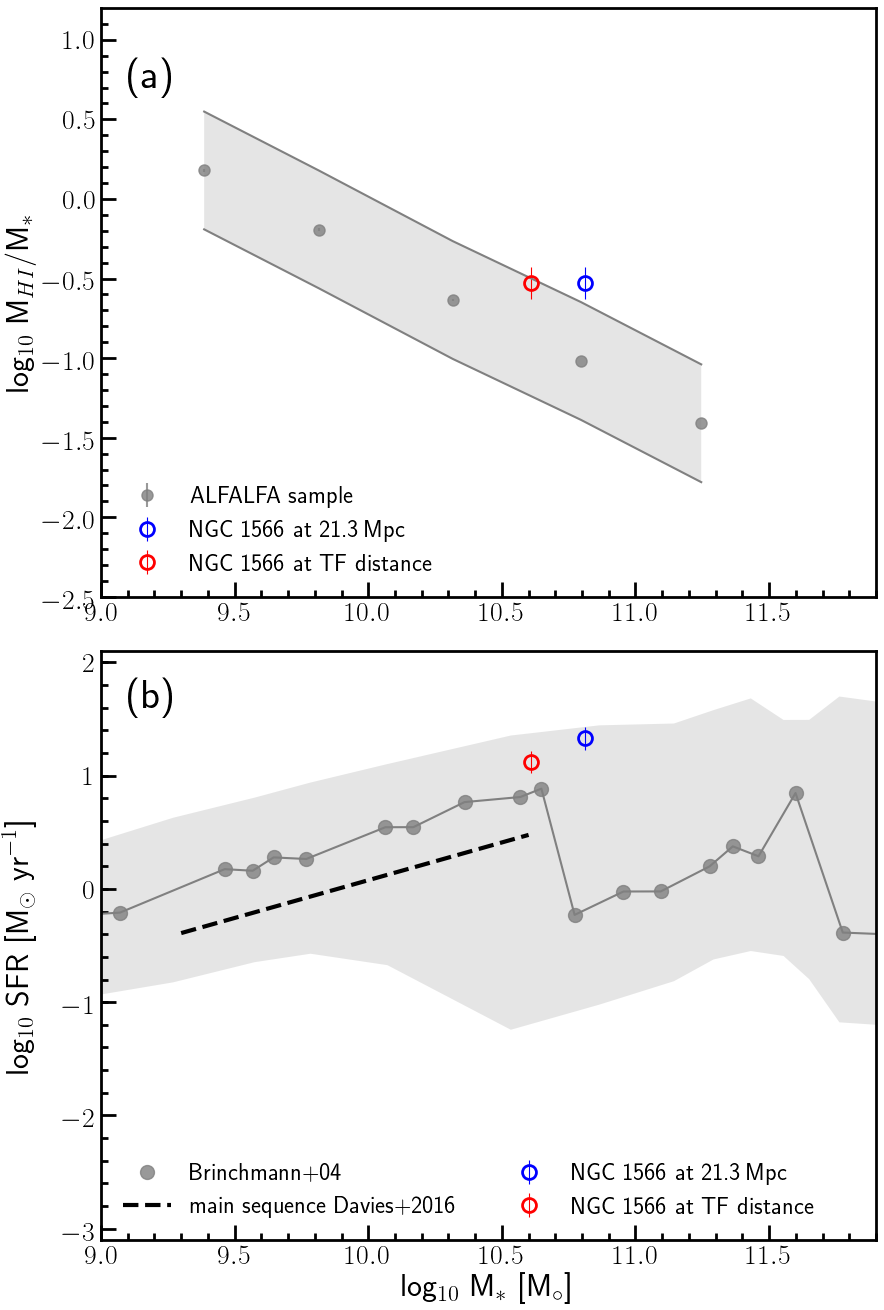}
\caption{(a)The H{\sc{i}}-to-stellar mass fraction vs. the stellar mass for Brown+15's galaxy
sample (grey) and for \mbox{NGC 1566} (blue circle). The red circle shows the location of \mbox{NGC 1566} in
the H{\sc{i}}-to-stellar mass fraction vs. M$_{*}$ plane measured at a distance of $16.9\,$Mpc (the TF distance).
The grey shaded regions delimit the y-axis scatter in  the  Brown+15's galaxy sample. (b) The star formation rate vs. stellar
mass for SDSS galaxies with $z < 0.2$;  the grey circles show the median value per stellar mass bin and 
the grey shaded region shows the scatter from the mean values along the SFR axis at a given stellar mass \citep{Brinchmann-2004}. The dashed-line shows the location of the main sequence of
star formation for the GAMA galaxy sample \citep{2016Davies}. The blue circle shows the location of \mbox{NGC 1566} in the SFR-M$_{*}$ plane. 
The red circle shows location of \mbox{NGC 1566} in the SFR-M$_{*}$ plane measured at the TF distance.}
\label{fig5_1}
\end{center}
\end{figure}

To demonstrate the high atomic gas content of NGC 1566, we compare the H{\sc{i}}-to-stellar mass fraction of this 
galaxy to a sample of $25,000$ galaxies within $10^{9}\,$M$_{\odot} \lesssim$ M$_{*}\lesssim10^{11.5}\,$M$_{\odot}$
and $0.02\leq z \leq0.05$ obtained from the Sloan Digital Sky Survey. \citet{Brown-2015} reported the
H{\sc{i}}-to-stellar mass fractions of these galaxies using  H{\sc{i}} data from the Arecibo Legacy Fast (ALFA) survey \citep{Giovanelli-2005}.
The stellar mass of \mbox{NGC 1566} is $M_{*}\,=\,6.5\times10^{10}\,$M$_{\odot}$, and has 
H{\sc{i}}-to-stellar mass fraction of log(M$_{\x{H{\sc{i}}}}$/M$_{*}$)$\,=\,-0.53\pm0.10$.
Figure \ref{fig5_1}a presents the H{\sc{i}}-to-stellar mass fraction vs. the stellar mass for 
Brown+15's sample and for \mbox{NGC 1566} (blue circle). It is evident that \mbox{\mbox{NGC 1566}} has a relatively 
high H{\sc{i}}-to-stellar mass fraction in comparison with its counterparts that have a stellar
mass of M$_{*}$ $\sim10^{10.81}\,$M$_{\odot}$. The average logarithmic H{\sc{i}}-to-stellar mass fraction of 
galaxies with M$_{*}$ $\sim10^{10.81}\,$M$_{\odot}$ is $\,-0.97$. We note that the uncertainty in the H{\sc{i}}-to-stellar mass of \mbox{NGC 1566} is within the upper bounds
of the scatter relation. This galaxy continues to have a relatively higher H{\sc{i}}-to-stellar mass fraction 
than the average at fixed stellar mass even when a smaller distance ($16.9\,$Mpc; the TF distance) is adopted (red circle in Figure \ref{fig5_1}a).
Even though, the outskirts of \mbox{NGC 1566} maybe subjected to ram-pressure 
interaction with the IGM, this scenario is not inconsistent with a high atomic gas fraction. Ram pressure 
stripping in group environments is likely to be subtle in comparison with galaxy clusters \citep{Westmeier-2011}, in which the density 
of the intra-cluster medium is orders of magnitude higher than in the IGM \citep{1998Eke,2009Pratt,2017Barnes}.\\

\begin{figure}
\begin{center}
\includegraphics[width=0.48\textwidth]{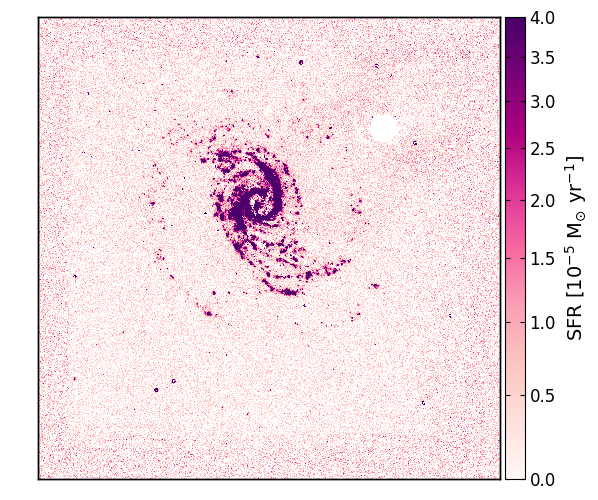}
\caption{The SFR map based on the H$\alpha$ luminosity of \mbox{NGC 1566} calculated following 
the SFR calibration of \citet{Meurer2009}. The star formation
is mainly concentrated in the inner spiral arms with some large H{\sc{ii}} regions distributed in the outer arms.}
\label{fig5_2}
\end{center}
\end{figure}

Figure \ref{fig5_1}b shows the relation between the star formation rate and stellar mass for $\sim10^5$ galaxies in the nearby Universe
($z < 0.2$) observed in the Sloan Digital Sky Survey (SDSS) and reported in \citet{Brinchmann-2004}. 
The dashed-line shows the location of the main sequence of star formation for over $10^5$ galaxies  in the local Universe (median redshift $z\,=\,0.2$) 
obtained by the Galaxy And Mass Assembly (GAMA) survey and reported in \citet{2016Davies}. The blue circle shows the star formation rate of \mbox{NGC 1566}, 
log(SFR$_{\x{H}\alpha}$[M$_{\odot}$ yr$^{-1}$])$\,=\,1.33\pm0.1$. This figure highlights the high SFR in \mbox{NGC 1566}. The star formation in this galaxy 
(Figure \ref{fig5_2}) is mainly concentrated in the nucleus and the inner spiral arms and declines gradually following the outer spiral
arms, in a similar fashion to the H{\sc{i}} gas. Further, this galaxy has a specific star formation rate (sSFR\,$=$\,log (SFR/M$_{*}$[yr$^{-1}$])) of 
\mbox{$-9.48\pm0.05$}, which again places \mbox{NGC 1566} above the average with respect
to galaxies that have the same stellar mass, but within the scatter \citep{2018Pan, Abramson-2014}. 
We note that the location of NGC 1566 in the mass-SFR parameter-space is subject to the  distance adopted for this galaxy.
For instance, if we use the TF distance ($16.9\,$Mpc) instead of the adopted distance in this work ($21.3\,$Mpc), the SFR and the stellar mass
of \mbox{NGC 1566} will be a factor of $0.63$ smaller. Thus, the SFR of NGC 1566 will become relatively closer to the median value 
per fixed stellar mass (red circle in Figure \ref{fig5_1}b), and NGC 1566 will appear less extreme in this parameter-space.\\

\section{Conclusion}

In this work, we present our ASKAP H{\sc{i}} observations of \mbox{NGC 1566}, a grand design spiral in the Dorado group.
Our major results and conclusions from this analysis are as follows:
\begin{itemize}
 \item We measure an H{\sc{i}} mass of 
$1.94\times10^{10}\,$M$_{\odot}$, assuming  a distance of $21.3\,$Mpc, mainly concentrated in the spiral arms 
of NGC 1566. The H{\sc{i}} gas is distributed in an  almost regular circular disc that extends well beyond the observed optical disc especially 
around the northern and western parts of \mbox{NGC 1566}. The H{\sc{i}} gas distribution also highlights the difference between 
the two outer arms better than in the optical. The eastern outer arm forms a regular arc shape that is more extended 
than the western arm, which is  less regular or disturbed between  {\textit{PA}}$\,=\,270^\circ$ to $330^\circ$.
The H{\sc{i}} disc of \mbox{NGC 1566} is asymmetric: the south-eastern part of the H{\sc{i}} disc
sharply declines beyond $\sim30\,$kpc from the centre, whereas the north-western
edge is more extended and smoothly declines with radius, up to a radius of $\sim50\,$kpc.\\

\item We measure the rotation curve of \mbox{NGC 1566} out to a radius of $50$ kpc and estimate the dark matter content 
in this galaxy based on the ISO, NFW and Burkert dark matter halo profiles. We report dark matter fractions 
of $0.66$, $0.58$ and $0.62$ based on the ISO, NFW and Burkert profiles, respectively. 
Using our current ASKAP observations, we can not differentiate between these dark matter density profiles as the central region ($\sim2\,$kpc) of \mbox{NGC 1566} is not resolved.
Higher angular resolution observations (few hundred parsec scales) are required for such analysis and for any conclusive findings \citep[see for example][]{2015seheon, 2008deBlok, 2002deBlok}. 
Such high angular resolution observations will be achieved in the next coming years using the $36$ ASKAP antennas 
(longest baseline of $6\,$km) and MeerKAT telescope \citep{2016deblok} for nearby galaxies. 
This will increase the number statistics of highly resolved rotation curves, 
and consequently better constraints on both the dark and baryonic matter distributions 
within these galaxies.\\

\item We study the asymmetric H{\sc{i}} morphology of \mbox{NGC 1566} and attempt to discriminate between  three major 
environmental mechanisms that can cause asymmetries in galaxies, 
namely, ram pressure interactions with the IGM, galactic interactions as well as gas accretion from  hosting/neighbouring
filaments. We detect no nearby companion galaxy that may induce 
tidal forces on the H{\sc{i}} disc of \mbox{NGC 1566} or tidal tails/plumes that are suggestive of such an encounter within the last $3.1\,$Gyr.
The H{\sc{i}} mass detection limit of the ASKAP observations based on the $3\sigma$ noise 
level and over $40\,$km s$^{-1}$ channel widths is $\sim2.2\times10^{7}\,$M$_{\odot}$ beam$^{-1}$ 
at the assumed distance of \mbox{NGC 1566} ($21.3\,$Mpc). We show, based on a simple  analytic model, that ram pressure stripping can affect the H{\sc{i}} disc of \mbox{NGC 1566} and is able to 
remove gas beyond a radius of $40$ kpc, using lower-limit values for the gas density of
the IGM and the relative velocity of this galaxy. Further, we do not detect any clouds or filaments connected to, or in the 
nearby vicinity of, the H{\sc{i}} disc of \mbox{NGC 1566}. However, we are unable to completely rule out gas accretion from the local environment at lower column densities. 
Future H{\sc{i}}  surveys with the SKA precursors and with large single dish telescopes, such as the
Five-hundred-meter Aperture Spherical radio Telescope \citep[FAST;][]{2011Nan, 2016Li, 2019Zhang}, will help  probe the environment around galaxies 
and quantify the prevalence of gas accretion, interactions and ram pressure stripping in large sample of galaxies and their effects on 
the atomic gas morphology and kinematics.\\

\item \mbox{\mbox{NGC 1566}} has a relatively high H{\sc{i}}-to-stellar mass fraction in
comparison with its counterparts that have the same stellar mass. The 
average logarithmic H{\sc{i}}-to-stellar mass fraction of galaxies with M$_{*}$ $\sim10^{10.81}\,$M$_{\odot}$ is
log(M$_{\x{H{\sc{i}}}}$/M$_{*}$)$=\,-0.97$. while for \mbox{NGC 1566} is log(M$_{\x{H{\sc{i}}}}$/M$_{*}$)$\,=\,-0.53\pm0.1$.
Further, \mbox{NGC 1566} possesses a specific star formation rate (sSFR\,$=$\,log SFR/M$_{*}$[yr$^{-1}$]) of 
\mbox{$-9.48\pm0.05$}, which is again above the average with respect to galaxies that have the same stellar mass, but within the 
scatter \citep{2018Pan, Abramson-2014}. However, the location of NGC 1566 in the mass-SFR 
parameter-space is dependent on the assumed distance, for which there remains significant uncertainty.\\

\end{itemize}

\section*{Acknowledgements}
We thank the anonymous referee for their positive and constructive comments which greatly improved the 
presentation of the results in this manuscript. AE is thankful for Davide Punzo and Kelley Hess for their help in making the 3D visualisation of \mbox{NGC 1566}, and for Kyle Oman for providing the
theoretical predictions for the relative velocity PDF of \mbox{NGC 1566} from his N-body simulations and libraries.
AB acknowledges financial support from the CNES (Centre National d'Etudes Spatiales, France). JW thank support from the National Science Foundation of China (grant 11721303).
This research was supported by the Australian Research Council Centre of Excellence for All-sky Astrophysics in 3 Dimensions (ASTRO 3D) through project number CE170100013.
The ATCA is part of the Australia Telescope National Facility (ATNF) and is operated by CSIRO. 
The ATNF receives funds from the Australian Government. This work was supported by 
resources provided by the Pawsey Supercomputing Centre with funding from the Australian Government
and the Government of Western Australia, including computational resources provided by the Australian
Government under the National Computational Merit Allocation Scheme (project JA3). PS has received funding from the European 
Research Council (ERC) under the European Union's Horizon 2020 research and innovation program (grant number
679629;name FORNAX). This paper used archival  H{\sc{i}}  data of \mbox{NGC 1566} available in the Australia Telescope Online Archive 
(http://atoa.atnf.csiro.au). ASKAP is part of the ATNF and is operated by CSIRO. The 
Operation of ASKAP is funded by the Australian Government with support from the National Collaborative Research Infrastructure Strategy.
ASKAP uses the resources of the Pawsey Supercomputing Centre. Establishment of ASKAP, the Murchison Radio-astronomy Observatory
and the Pawsey Supercomputing Centre are initiatives of the Australian Government, with support from the Government of Western
Australia and the Science and Industry Endowment Fund.
We acknowledge the Wajarri Yamatji people, the custodians of the observatory land. 
This work used images of \mbox{NGC 1566}  available in the NASA/IPAC \mbox{Extragalactic} Database (NED) and the Digitised Sky Surveys (DSS) website. 
NED is managed by the JPL (Caltech) under contract with NASA, whereas DSS is managed by the Space Telescope Science Institute (U.S. grant number NAG W-2166).
We also used infrared and ultraviolet images of \mbox{NGC 1566} from the Spitzer Space Telescope and the NASA Galaxy Evolution Explorer websites,
both space missions were managed by JPL under contract with NASA.

\bibliographystyle{mnras.bst}
\bibliography{774}

\appendix
\section{The Atomic Gas Morphology in \mbox{NGC 1566}}

\begin{figure*}
\includegraphics[width=0.9\textwidth]{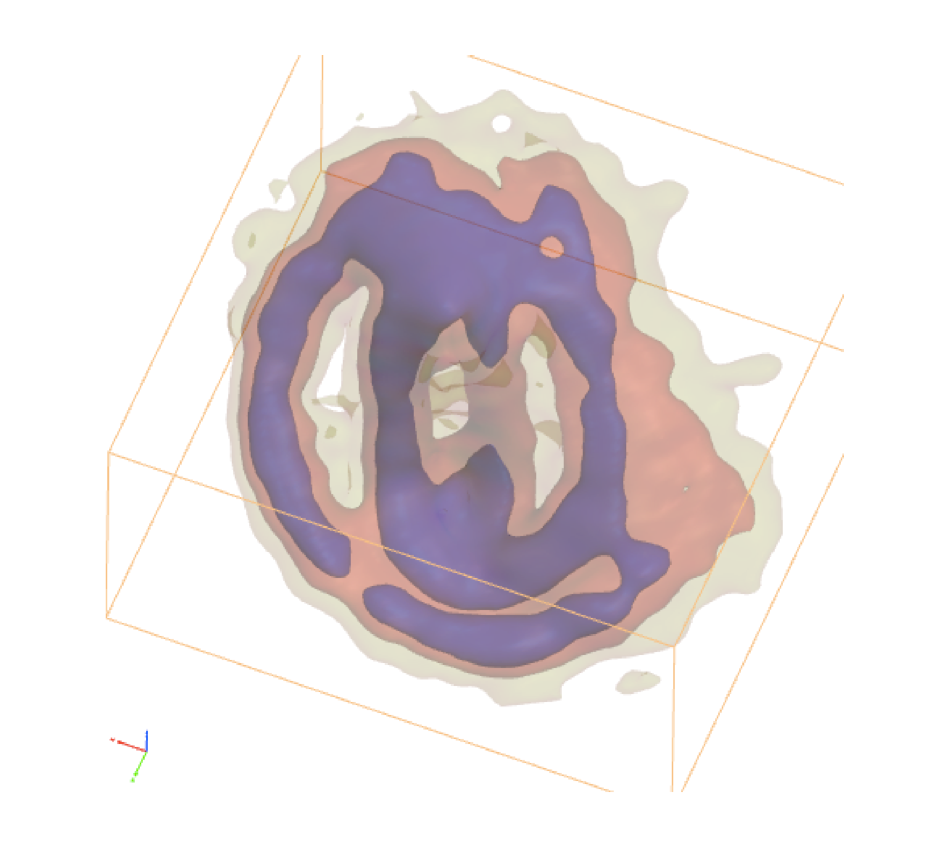}
\caption{3D rendering of the H{\sc{i}} cube of \mbox{NGC 1566} based on our ASKAP early science observations,
the three axes in this visualisation are RA, Dec, and the velocity . To improve the visualisation, pixels with flux values 
$> 0.088,  0.037\,$Jy beam$^{-1}$ km$^{-1}$ and $0.0017\,$Jy beam$^{-1}$ km$^{-1}$ are rendered with  blue, red and green colors, respectively.}
\label{3dplot}
\end{figure*}

Figure \ref{3dplot} shows a 3D position-position-velocity interactive view of the H{\sc{i}} cube of \mbox{NGC 1566} made
using the Astronomy H{\sc{i}} Extension for 3D Slicer \citep[SlicerAstro;][]{slicer1,slicer2}\footnote{\url{https://github.com/Punzo/SlicerAstro}}
\footnote{\url{https://www.slicer.org}}. The reader can actively interact (rotate, drag, zoom-in or out) with the 3D plot using Adobe Reader 9.

\label{lastpage}

\bsp

\end{document}